# X-RAY SOURCE POPULATIONS IN GALAXIES

G. Fabbiano

*Harvard-Smithsonian Center for Astrophysics*
*60 Garden St., Cambridge, Massachusetts 02138*
e-mail: gfabbiano@cfa.harvard.edu

**Abstract**
Today's sensitive, high-resolution X-ray observations allow the study of populations of X-ray sources, in the luminosity range of Galactic X-ray binaries, in galaxies as distant as 20-30Mpc. The traditional astronomical tools of photometric diagrams and luminosity functions are now applied to these populations, providing a direct probe of the evolved binary component of different stellar populations. The study of the X-ray populations of E and S0 galaxies has revamped the debate on the formation and evolution of low-mass X-ray binaries (LMXBs) and on the role of globular clusters in these processes. While overall stellar mass drives the amount of X-ray binaries in old stellar populations, the amount of sources in star forming galaxies is related to the star formation rate. Short-lived, luminous, high mass binaries (HMXBs) dominate these young X-ray populations. The most luminous sources in these systems are the debated ULXs, which have been suggested to be ~100-1000 $M_\odot$ black holes, but could alternatively include a number of binaries with stellar mass black holes. Very soft sources have also been discovered in many galaxies and their nature is currently being debated. Observations of the deep X-ray sky, and comparison with deep optical surveys are providing the first evidence of the X-ray evolution of galaxies.

## 1. *CHANDRA*: A NEW PARADIGM

This review comes almost two decades after the 1989 Annual Review article on the X-ray emission from galaxies (Fabbiano 1989), and a few words on the evolution of this field are in order. In 1989, the *Einstein Observatory* (Giacconi *et al.* 1979), the first imaging X-ray telescope, had opened up the systematic study of the X-ray emission of normal galaxies. The *Einstein* images, in the ~0.3 – 4 keV range, with resolutions of ~5'' and ~45'' (see the *Einstein Catalog and Atlas of Galaxies*, Fabbiano, Kim & Trinchieri 1992) showed extended and complex X-ray emission, and gave the first clear detection of individual luminous X-ray sources in nearby spiral galaxies, other than the Milky Way. The first ultra-luminous (non-nuclear) X-ray sources (ULXs) were discovered with *Einstein*, and the suggestion was advanced that these sources may host >100 $M_\odot$ black holes, a topic still intensely debated. Hot diffuse halos were discovered in elliptical galaxies and used as a means of estimating the mass of the dark matter associated with these galaxies, but their ubiquity and properties were hotly debated. Super-winds from actively star-forming galaxies (e.g. M82), an important component of the ecology of the universe, were first discovered with *Einstein*. All these topics are discussed in the 1989 review.

The subsequent X-ray observatories *ROSAT* (Truemper 1983) and *ASCA* (Tanaka *et al.* 1994) expanded our knowledge of the X-ray properties of galaxies (see e.g., a review summary in Fabbiano & Kessler 2001), but did not produce the revolutionary leap originated by the first *Einstein* observations. The angular resolution of these missions was comparable (*ROSAT*) or inferior (*ASCA*, with 2 arcmin resolution) to that of Einstein, but the *ROSAT* spectral band (extending down to ~0.1 keV) and lower background, provided a better view in some cases of the cooler X-ray components (halos and hot outflows), while the wide spectral band (~0.5 – 10 keV) and better spectral resolution of the *ASCA* CCD detectors allowed both the detection of emission lines in these hot plasmas and the spectral decomposition of integrated emission components (e.g., Matsushita *et al.* 1994). Overall, however, many of the questions raised by the *Einstein* discoveries remained (see Fabbiano & Kessler 2001). It is only with *Chandra*'s sub-arcsecond angular resolution (Weisskopf *et al.* 2000), combined with photometric capabilities commensurable with those of *ASCA*, that the study of normal galaxies in X-rays has taken a second revolutionary leap. With *Chandra*, populations of individual X-ray sources, with luminosities comparable to those of Galactic X-ray binaries, can be detected at the distance of the Virgo Cluster and beyond; the emission of these sources can be separated from the diffuse emission of hot interstellar gases, both spatially and spectrally; detailed measures of the metal abundance of these gaseous components can be attempted (e.g., Soria & Wu 2002; Martin, Kobulnicky & Heckman 2002; Fabbiano *et al.* 2004a; Baldi *et al.* 2005a, b); and quiescent super-massive nuclear black holes can be studied (e.g. Fabbiano *et al.* 2004b; Pellegrini 2005; Soria *et al.* 2005).

Here, I will concentrate only on one aspect of the emission of normal galaxies, the study of their X-ray source populations, avoiding detailed discussions of the properties of individual nearby galaxies (see Fabbiano & White 2005 for an earlier review of this topic, based on publications up to 2003). Also, I will not discuss the properties of the hot interstellar medium (ISM) and of low-level nuclear emission, which were all included in the 1989 review. The field has expanded enough since then, that these topics now deserve their separate reviews. Most of the work discussed in this review is the result of the study of high resolution *Chandra* images. Whenever relevant, (for the most nearby galaxies, and the spectral study of ULXs), I will also discuss observations with *XMM-Newton* (the *European Space Agency* X-ray telescope, with an effective area ~3 times larger than *Chandra*, but significantly coarser angular resolution ~15'').

The plan of this paper is as follows: Section 2. is a short discussion of the observational and analysis approaches opened by the availability of high resolution, sensitive X-ray data; Section 3. reviews the results on the old X-ray binary population found in early-type galaxies and spiral bulges; Section 4. addresses the work on the younger stellar population of spiral and irregular galaxies; Sections 5.

and 6. discuss two classes of rare X-ray sources, to the understanding of which recent observations of many galaxies have contributed significantly: super-soft sources (SSSs) and ULXs; Section 7. concludes this review with a short discussion of the properties of the galaxies observed in deep X-ray surveys.

## 2. POPULATION STUDIES IN X-RAYS

It is well known that the Milky Way hosts both old and young luminous X-ray source populations, reflecting its general stellar make up. In the luminosity range detectable in most external galaxies with typical *Chandra* observations ( $> 10^{37}$ erg s$^{-1}$), these sources are prevalently X-ray binaries (XRBs). Old population Galactic X-ray sources are accreting neutron star or black hole binaries with a low-mass stellar companion (LMXBs, with life-times ~$10^{8-9}$ yr). Young population X-ray sources, in the same luminosity range, are dominated by neutron star or black hole binaries with a massive stellar companion (HMXBs, with life-times ~ $10^{6-7}$ yr), although a few young supernova remnants (SNRs) may also be expected. At lower luminosities, reachable with *Chandra* in Local Group galaxies, Galactic sources include accreting white dwarfs and more evolved SNRs (see e.g., the review by Watson 1990; Grimm, Gilfanov & Sunyaev 2002 for a study of the X-ray luminosity functions of the Galactic X-ray source populations; see Verbunt & van den Heuvel 1995 for a review on the formation and evolution of X-ray binaries). Fig.1 shows the cumulative X-ray luminosity functions (XLFs) of LMXBs and HMXBs in the Galaxy (Grimm, Gilfanov & Sunyaev 2002). Note the high luminosity cut-off of the LMXB XLF and the power-law distribution of the HMXB XLF; these basic characteristics are echoed in the XRB populations of external galaxies (Sections 3.3 and 4.2).

Fig. 2 shows two typical observations of galaxies with *Chandra*: the spiral M83 (Soria & Wu 2003) and the elliptical NGC4697 (Sarazin, Irwin & Bregman 2000), both observed with the ACIS CCD detector. The images are color coded to indicate the energy of the detected photons (red 0.3-1 keV, green 1-2 keV and blue 2-8 keV). Populations of point-like sources are easily detected above a generally cooler diffuse emission from the hot interstellar medium. Note that luminous X-ray sources are relatively sparse by comparison with the underlying stellar population, and can be detected individually with the *Chandra* sub-arcsecond resolution, with the exception of those in crowded circum-nuclear regions.

The X-ray CCD detectors (present both in *Chandra* and *XMM-Newton*) provide us with a data-tesseract of the observed area of the sky, where each individually detected photon is tagged with a 2-dimensional position, an energy and a time of arrival. So, for each detected source, we can measure its flux (and luminosity), have some sort of spectral (or photometric) information and variability as well. For the most intense sources, it is also possible to derive correlated variability-spectral information, if the galaxy has been observed at different epochs (which it is still rare in the available data set; see, e.g. Fabbiano *et al.* 2003a, b).

To analyze this wealth of data two approaches have been taken: (1) a photometric approach, consisting of X-ray color-color diagrams and color-luminosity diagrams, and (2) X-ray luminosity functions.

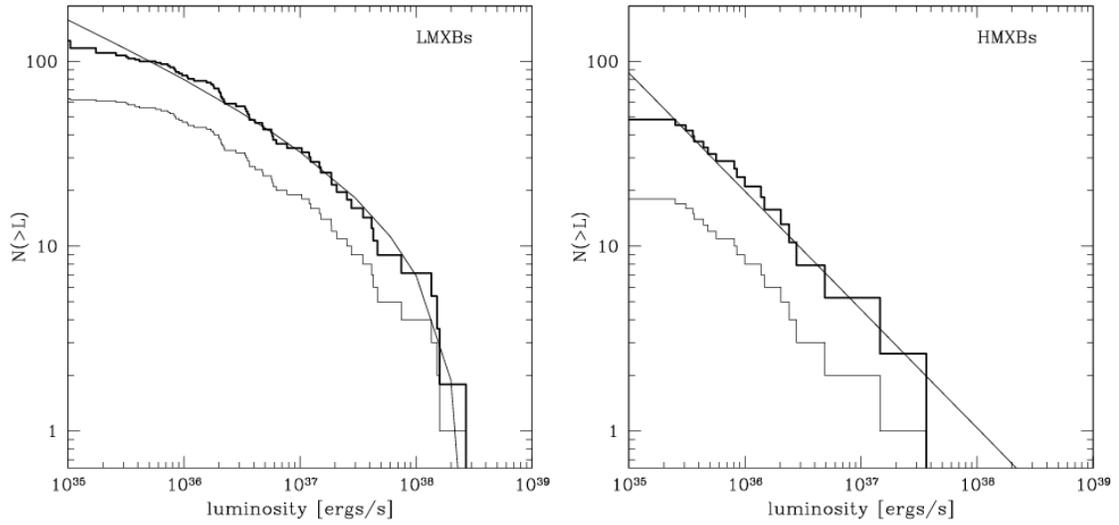

Fig. 1 –Cumulative XLFs of Galactic LMXBs (Left) and HMXBs (right), from Grimm, Gilfanov and Sunyaev (2002).

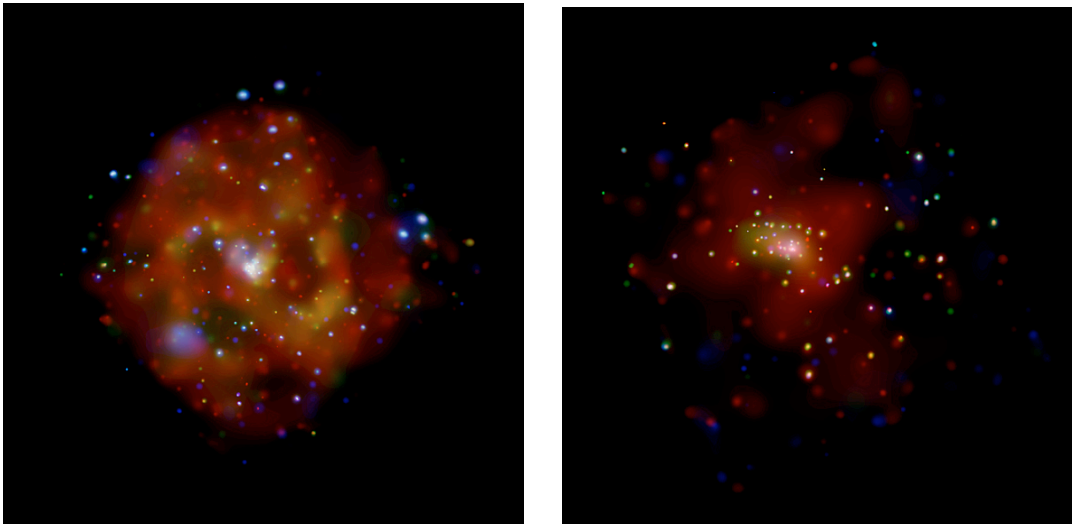

Fig.2 – Chandra ACIS images of M83 (left, box is 8.57 × 8.86 arcmin) and NGC4697 (right, box is 8.64 × 8.88 arcmin). See text for details. Both images are from the web page http://chandra.harvard.edu/photo/category/galaxies.html; credit NASA/CXC).

## 2.1 X-ray photometry

The use of X-ray colors to classify X-ray sources is not new. For example, White & Marshall (1984) used this approach to classify Galactic XRBs, and Kim, Fabbiano &

Trinchieri (1992) used *Einstein* X-ray colors to study the integrated X-ray emission of galaxies. Unfortunately, given the lack of standard X-ray photometry to date, different definitions of X-ray colors have been used in different works; in the absence of instrument corrections, these colors can only be used for comparing data obtained with the same observational set up. Colors, however, have the advantage of providing a spectral classification tool when only a limited number of photons are detected from a given source, which is certainly the case for most X-ray population studies in galaxies.

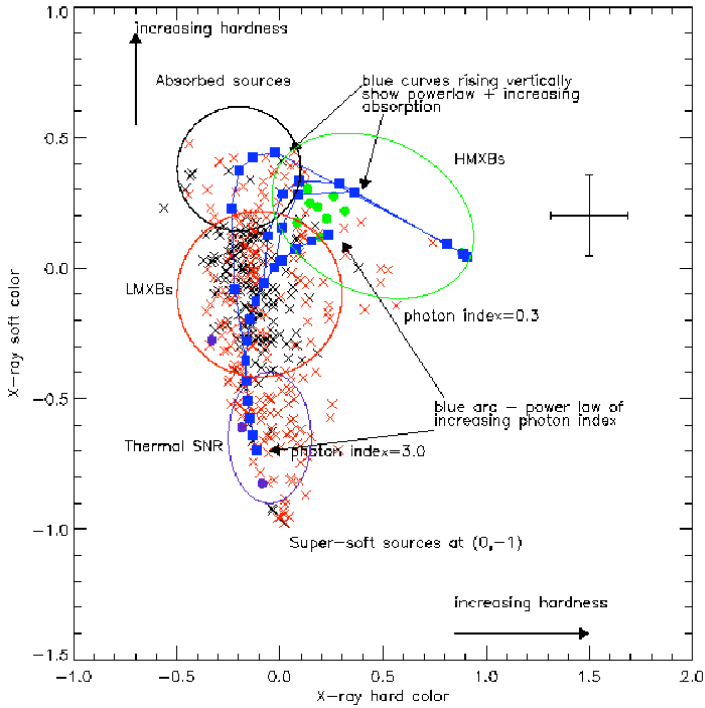

Fig. 3 – Chandra color-color diagram from Prestwich *et al.* (2003)

Also, compared with the traditional derivation of spectral parameters via model fitting, color-color diagrams provide a relatively assumption-free comparison tool. *Chandra*-based examples of this approach can be found in Zezas *et al.* 2002a, b and Prestwich *et al.* 2003, among others. The X-ray color-color diagram of Prestwich *et al.* 2003 (fig. 3) illustrates how colors offer a way to discriminate among different types of possible X-ray sources.

## 2.2 X-ray luminosity Functions (XLFs)

Luminosity functions are a well known tool in observational astrophysics. XLFs have been used to characterize different XRB populations in the Milky Way (e.g., Grimm, Gilfanov & Sunyaev 2002, see Fig. 1), but these studies have always required a model of the spatial distribution of the sources, in order to estimate their luminosities, which is inherently a source of uncertainty. External galaxies, instead, provide clean source samples, all at the same distance. Moreover, the detection of X-ray source populations in a wide range of different galaxies allows us to explore global population differences that may be connected with the age and or metallicity of the parent stellar populations. XLFs establish the observational basis of X-ray population synthesis (Belczinsky *et al.* 2004).

While in principle XLFs are simple to construct, they do suffer from observational biases and statistical effects, which must be either corrected or accounted for. The incomplete detection of low-luminosity sources is a well-known effect that may cause flattening of the XLF at the low luminosity end. The artificial `brightening' of threshold sources because of statistical fluctuations is another well-known effect (Eddington bias). To these, we must add the varying amount of diffuse emission around the source (from a hot interstellar medium, present in varying quantities and with both spatial and spectral variations in galaxies, e.g., Zezas & Fabbiano 2002), which also affects the detection threshold, and source confusion in crowded regions especially near the galaxy centers. In the case of *Chandra* we also need to consider the radial dependence of the degradation of the mirror resolution off-axis, (see Kim & Fabbiano 2003; Kim & Fabbiano 2004; Gilfanov 2004). These low-luminosity biases have not been treated consistently in the literature, giving rise in some cases to potentially spurious results (Section 3.3.1). For galaxies extending over large angular sizes, the effect of background AGNs and stellar interlopers in the XLF should also be considered (e.g. Finoguenov & Jones 2002; Gilfanov 2004; Grimm *et al.* 2005).

The evaluation of the total X-ray luminosity of a galaxy may be significantly affected by statistics when a relatively small number of X-ray sources are detected (Gilfanov, Grimm & Sunyaev 2004b). In particular, the paucity of very luminous X-ray sources in galaxies makes uncertain the definition of the high luminosity XLF, which may be better approached by coadding `consistent' samples of X-ray sources (e.g., Kim & Fabbiano 2004).

Compact X-ray sources are notorious for their variability and this variability could in principle also affect the XLF, which is typically derived from a snapshot of a given galaxy. However, repeated *Chandra* observations in the case of NGC5128 (Kraft *et al.* 2001), M33 (Grimm *et al.* 2005) and the Antennae galaxies (Zezas *et al.* 2004) empirically demonstrate that the XLF is remarkably steady against individual source variability.

## 3. OLD XRB POPULATIONS

At variance with most previous reviews of X-ray observations of galaxies, which tend to concentrate first on nearby well studied spiral and irregular galaxies, I will begin by discussing the X-ray populations of old stellar systems: E and S0 galaxies. By comparison with spirals, these galaxies present fairly homogeneous stellar populations, and therefore one can assume that their XRB populations are also more uniform, providing a 'cleaner' baseline for population studies.

### 3.1 LMXBs in Early-type galaxies: there they are – Past and present

The 1989 review (Fabbiano 1989) discussed the presence and in some case predominance of LMXB X-ray emission in E and S0 galaxies. This was a controversial issue at the time, because LMXBs could not be detected individually, and their presence was supported only by statistical considerations (e.g., Trinchieri & Fabbiano 1985). Although the spectral signature of LMXBs was eventually detected (Kim, Fabbiano & Trinchieri 1992; Fabbiano, Kim & Trinchieri 1994; Matsushita *et al.* 1994), uncontroversial detection of samples of these sources in all early type galaxies has become possible only with the sub-arcsecond resolution of *Chandra* (such a population was first reported in NGC 4697, where 80 sources were detected by Sarazin, Irwin & Bregman 2000, see Fig. 1).

A statistical analysis of a large sample of early-type galaxies observed with *Chandra* is still to come, but the results so far confirm the early conclusion (see Fabbiano 1989; Kim, Fabbiano & Trinchieri 1992; Eskridge, Fabbiano & Kim 1995a, b) that LMXBs can account for a very large fraction of the X-ray emission of some early type galaxies (those formerly known as 'X-ray faint', i.e. devoid of large hot gaseous halos): for example, in NGC4697 (Sarazin, Irwin & Bregman 2000) and NGC1316 (Kim & Fabbiano 2003) the fraction of detected counts attributable to the hot ISM is ~23% and ~50% respectively. In both cases, given the harder spectrum of LMXBs, these sources dominate the total luminosity of the galaxy in the $0.3 - 8$ keV range. In NGC1316 the integrated LMXB emission, including non-detected LMXBs with luminosities below threshold, could be as high as $4 \times 10^{40}$ erg s$^{-1}$. Sivakoff, Sarazin and Irwin (2003) reach similar conclusions for NGC4365 and NGC4382.

Although this review focuses on the X-ray binary populations, I cannot help remarking that the *Chandra* results demonstrate unequivocally that ignoring the contribution of the hidden emission of LMXBs was clearly a source of error in past estimates of galaxy dynamical mass, as discussed in the 1989 review (see also Trinchieri, Fabbiano & Canizares 1986). NGC1316 (Kim & Fabbiano 2003) provides a very clear illustration of this point. In this galaxy the LMXBs are distributed like the optical light, and dominate the emission at large radii. Instead, the ISM follows a steeper profile (Fig 4 Left), with temperature possibly decreasing at larger radii, suggestive of winds. Use of lower resolution *Einstein* data, with the assumption of hydrostatic equilibrium, resulted in a large mass estimate for this galaxy (Forman, Jones & Tucker 1985), which is clearly not sustained by the present data, since the extent of the gaseous halo is less than assumed, its temperature is lower (because the *Einstein* spectrum was clearly contaminated by the harder LMXBs emission), and the halo may not be in hydrostatic equilibrium. Similarly, spectral analysis of the NGC1316 *Chandra* data, after subtraction of the detected LMXBs, and taking into account the unresolved LMXB component, results in constraints on the metallicity of the hot ISM ($Z = 0.25-1.3\ Z_\odot$) more in keeping with the values expected from stellar evolution (e.g. Arimoto *et al.* 1997) than the extremely sub-solar values (0.1 solar, Iyomoto *et al.* 1998, in the case of NGC1316), typically resulting from *ASCA* data analysis of the integrated emission of the whole galaxy.

## 3.2 Source spectra and variability

Populations of several tens to hundreds sources have been detected in a number of E and S0 galaxies with *Chandra* (see review by Fabbiano & White 2005), and the number is growing as both the number of galaxies observed and the depth of the observations increase. With the exception of a few super-soft sources reported in some galaxies (see Irwin, Athey & Bregman 2003; Humphrey & Buote 2004), the X-ray colors and spectra of these sources are consistent with those expected of LMXBs, and consistent with those of the LMXBs of M31 (Sarazin, Irwin & Bregman 2001; Blanton, Sarazin & Irwin 2001; Finoguenov & Jones 2002; Kim & Fabbiano 2003; Irwin, Athey & Bregman 2003; Sivakoff, Sarazin & Irwin 2003; Kim & Fabbiano 2004; Randall, Sarazin & Irwin 2004; Humphrey & Buote 2004; Jordan et 2004; Trudolyubov & Priedhorsky 2004; David *et al.* 2005).

The most extensive spectral study to date is that of Irwin, Athey & Bregman (2003), who studied 15 nearby early-type galaxies observed with *Chandra*. These authors found that the average spectrum of sources fainter than $10^{39}$ erg s$^{-1}$ is remarkably consistent from galaxy to galaxy, irrespective of the distance of the sources from the center of the galaxy. These spectra can be fitted with either power-laws with photon index $\Gamma=1.56 \pm 0.02$ (90%) or with bremmstrahlung emission with kT=7.3 ± 0.3 keV. Sources with luminosities in the $(1 - 2) \times 10^{39}$ erg s$^{-1}$ range instead have softer spectra, with power-law $\Gamma \sim 2$, consistent with the high-soft emission of black hole binaries (mass of up to $15 M_\odot$ expected). Within the errors, these results are consistent with those reported in other studies, although sources in different luminosity ranges are usually not studied separately in these works. Jordan *et al.* (2004) confirm the luminosity dependence of the average source spectrum in M87; their color-color diagram suggests a spectral softening for sources more luminous than $5 \times 10^{38}$ erg s$^{-1}$.

Relatively little is known about source variability, since repeated monitoring of the same galaxy is not generally available. However, variable sources and at least five transients (dimming factor of at least 10) have been detected in NGC5128, with two *Chandra* observations (Kraft *et al.* 2001). Variable sources are also detected with two observations of NGC1399, taken two years apart (Loewenstein, Angelini & Mushotzky 2005). Sivakoff. Sarazin & Irwin (2003) report variability in a few sources in NGC 4365 and NGC4382 within 40ks *Chandra* observations; Humphrey & Buote (2004) report two variable sources in NGC1332. Sivakoff, Sarazin & Jordan (2005) report short time-scale X-ray flares from 3 out of 157 sources detected in NGC4697; two of these flares occurs in globular cluster sources and are reminiscent of the superbursts found in Galactic neutron star binaries, the third could originate from a black hole binary. Maccarone (2005) suggests that these flares may be periodical events resulting from periastron accretion of eccentric binaries in dense globular clusters.

The spectral characteristics of the point sources detected in E and S0 galaxies, their luminosities, and their variability, confirm the association of these sources with compact accreting X-ray binaries.

## 3.3 X-ray Luminosity Functions (XLFs) of LMXB populations

The luminosities of individual detected sources range typically from a few $10^{37}$ erg s$^{-1}$, depending on the distance of the galaxy and the observing time, up to $\sim 2 \times 10^{39}$ erg s$^{-1}$. XLFs have been derived in most *Chandra* studies of early-type galaxies, to study the luminosity distributions of the LMXB populations. These XLFs (both in differential and cumulative forms) have been modeled to characterize their functional shape, infer the presence of breaks, and estimate the total LMXB contribution to the X-ray emission of the galaxies. In the following I first review the work on the high luminosity shape of the XLF and the possible presence of a break near the Eddington luminosity of an accreting neutron star. I then discuss the much less studied low-luminosity shape of the XLF, and finally address the characteristics and drivers of the normalization (i.e. the total LMXB content of a galaxy).

3.3.1 High luminosity shape ($L_X >$ a few $10^{37}$ erg s$^{-1}$)

The shape of the XLF has been parameterized with models consisting of power-laws or broken power-laws. The overall shape (in a single power-law approximation in the range of ~ $7 \times 10^{37}$ to a few $10^{39}$ erg s$^{-1}$), is fairly steep, i.e. with a relative dearth of high luminosity sources, when compared with the XLFs of star-forming galaxies (Section 4.2; see also Kilgard *et al.* 2002; Colbert *et al.* 2004; Fabbiano & White 2005), but the details of these shapes and the related presence of breaks have been matter of some controversy.

Two breaks have been reported in the XLFs of E and S0 galaxies: the first is a break at ~2-5 x $10^{38}$ erg s$^{-1}$, near the Eddington limit of an accreting neutron star, first reported by Sarazin, Irwin & Bregman (2000) in NGC4697, which may be related to the transition in the XLF between neutron star and black hole binaries (Blanton, Sarazin & Irwin 2001 in NGC1553; Finoguenov and Jones 2002 in M84; Kundu, Maccarone & Zepf 2002 in NGC4472; Jordan *et al.* 2004 in M87; Kim & Fabbiano 2004; Gilfanov 2004; see also Di Stefano *et al.* 2003 for the XLF of the Sa Sombrero galaxy, NGC4594); the second is a high luminosity break at ~$10^{39}$ erg s$^{-1}$, first discussed by Jeltema *et al.* (2003) in NGC720 (see also Sivakoff *et al.* 2003; Jordan *et al.* 2004). Both breaks are somewhat controversial, because the interpretation of the observed XLFs is crucially dependent on a proper completeness correction.

Incompleteness effects are particularly relevant for the Eddington break, because the typical exposure times of the data and the distances of the target galaxies in most cases conspire to produce a spurious break at just this value (as demonstrated for NGC1316 by Kim & Fabbiano 2003, fig. 4 Right). Interestingly, no break was required in the case of NGC5128 (Kraft *et al.* 2001), where the proximity of this galaxy rules out incompleteness near the neutron star Eddington luminosity. An apparent Eddington break that disappears after completeness correction is also found by Humphrey & Buote (2004) for the XLF of NGC1332. Similarly, Eddington breaks are absent in NGC4365 and NGC4382 (Sivakoff, Sarazin & Irwin 2003), while a high luminosity cut-off at 0.9-3.1 x $10^{39}$ erg s$^{-1}$ could be allowed; these authors also consider the effect of incompleteness in their results.

Other recent papers, however, do not discuss, or do not apply, completeness corrections to the XLFs, so their conclusions on the presence of Eddington breaks need to be confirmed. Randall, Sarazin & Irwin (2004) report a break at ~$5 \times 10^{38}$ erg s$^{-1}$ in NGC 4649, with large uncertainties, but do not discuss the derivation of the XLF. Jordan *et al.* (2004) derive and fit the XLF of M87, and compare it with their own fit of those of NGC4697 and M49 (NGC4472), using the data from Sarazin, Irwin & Bregman (2001) and Kundu, Maccarone & Zepf (2002) respectively. However, completeness corrections are not applied, although the low luminosity data are not fitted. Jordan *et al.* (2004) report breaks at $2\text{-}3 \times 10^{38}$ erg s$^{-1}$ in all cases, or a good fit with a single power law truncated at $10^{39}$ erg s$^{-1}$. Note that these results are not consistent with those of Kim & Fabbiano (2004) where the corrected XLFs of NGC4647 and NGC4472 are well fitted with single unbroken power-laws.

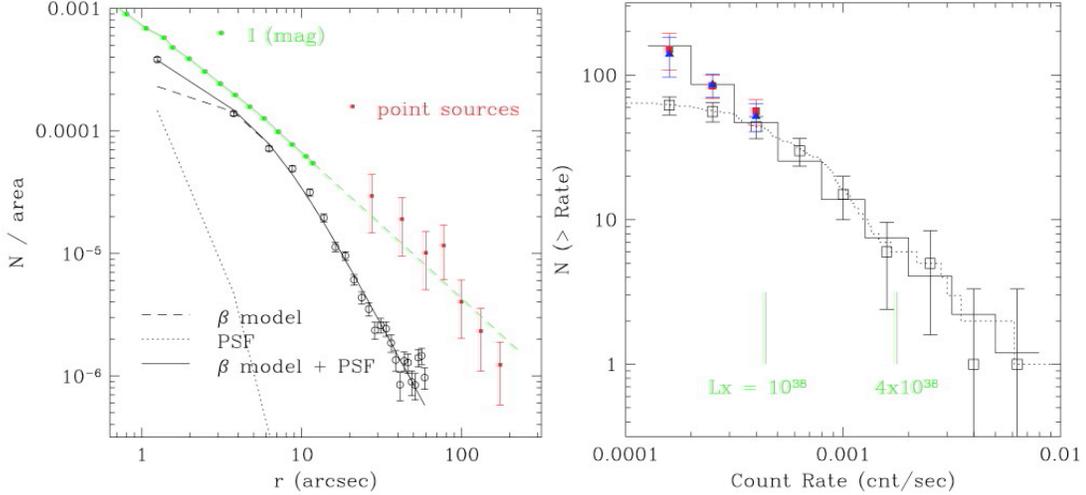

Fig. 4 – Left, radial profile of LMXBs of NGC1316 compared with optical light (green) and diffuse hot ISM emission (black); Right, XLF before and after completeness correction (both from Kim & Fabbiano 2003).

Kim & Fabbiano (2004) derive corrected luminosity functions for a sample of 14 E and S0 galaxies, including some with previously reported breaks, and find that all the individual corrected XLFs are well fitted with single power-laws with similar differential slopes (1.8–2.2) in the observed luminosity range. None of these fits require an Eddington break. However, a break may be hidden by the poor statistics in each case. The statistical consistency of the individual power-laws justifies co-adding the data to obtain a high significance composite XLF of early-type galaxies (fig. 5 Left). This composite XLF is not consistent with a single power-law, suggesting a break at $(5 \pm 1.6) \times 10^{38}$ erg s$^{-1}$. The best-fit differential slope is $1.8 \pm 0.2$ in the few $10^{37}$ to $5 \times 10^{38}$ erg s$^{-1}$ luminosity range for the co-added XLF; at higher luminosity, above the break, the differential slope is steeper ($2.8 \pm 0.6$). These results are confirmed by the independent work of Gilfanov (2004), who analyzes four early-type galaxies, included in the Kim & Fabbiano (2004) sample (fig. 5 Right); however, Gilfanov's differential slope for the high luminosity portion of the XLF is somewhat steeper (3.9 – 7.3). Both the Kim & Fabbiano and Gilfanov analyses are consistent with a cut-off of the XLF of LMXBs at a few $10^{39}$ erg s$^{-1}$. A more recent paper (Xu *et al.* 2005) is in agreement with the above conclusions, reporting a consistent break in the corrected XLF of NGC4552; on the basis of a simulation, this paper concludes that the break may or may not be detected in any individual galaxy XLF, given the relatively small number of sources present in each individual galaxy.

The $(5 \pm 1.6) \times 10^{38}$ erg s$^{-1}$ break is at somewhat higher luminosity that it would be expected for an Eddington break of normal neutron star binaries. It may be consistent with the luminosity of the most massive neutron stars ($3.2 \pm 1$ M$_\odot$; see Ivanova & Kalogera 2005), He-enriched neutron star binaries ($1.9 \pm 0.6$ M$_\odot$; see Ivanova & Kalogera 2005) or low-mass black hole binaries. This effect may be related to different populations of LMXBs in early type galaxies, including both neutron star and black hole binaries; it may also be the consequence of a true high luminosity break in the XLF (e.g. Sivakoff, Sarazin & Irwin 2003). Whatever the cause, the shape of the XLF points to a dearth of very luminous sources in E and S0 galaxies. Note that a high luminosity cut-off is also present in the XLF of Galactic LMXBs (fig. 1).

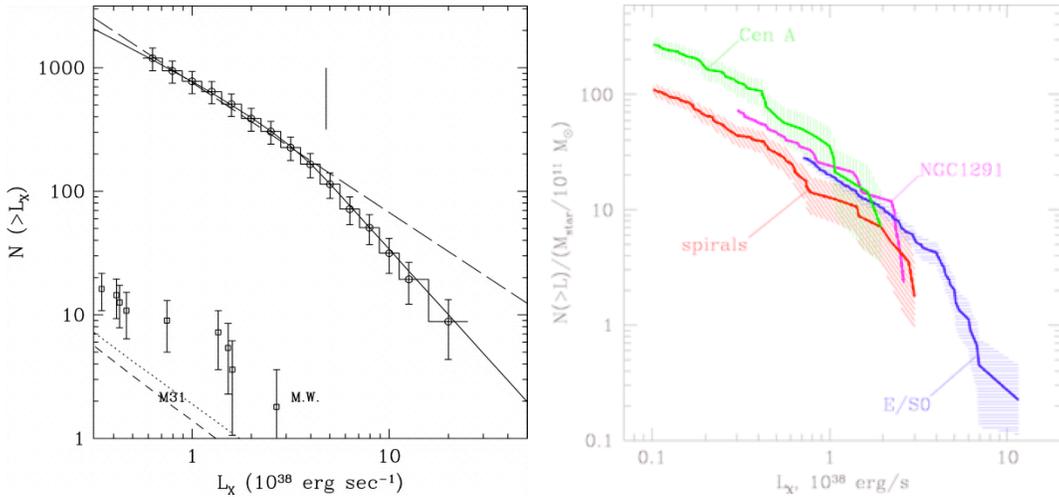

Fig. 5 – Left, cumulative XLF of 14 E and S0 galaxies (Kim & Fabbiano 2004), with the single power-law best fit (dashed), and the broken power-law model (solid line); the M31 and Milky Way LMXB XLFs are sketched in the left lower corner. Right, Cumulative LMXB XLFs from Gilfanov (2004). Note the similarity of the XLFs and the break at $\sim 5 \times 10^{38}$ erg s$^{-1}$ in the E/S0 XLF.

3.3.3 Low luminosity shape

With the exception of NGC5128 (Cen A), with a measured XLF extending down to $\sim 2 \times 10^{36}$ erg s$^{-1}$ (Kraft *et al.* 2001), the *Chandra* data presently available for E and S0 galaxies does not allow the detection of LMXBs with luminosities below the mid or high $10^{37}$ erg s$^{-1}$ range. By including Cen A and the LMXB (bulge) population of nearby spirals (Milky Way, M31, M81) in his study, Gilfanov (2004) suggests that the XLF flattens at lower luminosity, following differential slopes near 1 below $10^{37}$ erg s$^{-1}$ (fig. 5 Right). A recent reanalysis of the Cen A data confirms this result (Voss

& Gilfanov 2005). Direct deep observations of 'normal' early-type galaxies with *Chandra* are needed to see if this suggestion is generally valid. Moreover, detailed studies may show a more complex situation in the low luminosity XLFs. For example, in M31, the best-studied LMXB bulge population, clear differences are reported between the XLFs of bulge and globular cluster (GC) sources, and different low-luminosity breaks are present in the inner and outer bulge, and in GC XLFs (Kong *et al.* 2002; Kong *et al.* 2003; see discussion in review by Fabbiano & White 2005). The physical origins of these striking differences are not yet well understood, although it has been suggested that the radial dependence of the XLF break in the bulge of M31 could be related to an older population at the inner radii (Kong *et al.* 2002). Certainly, these features suggest an unexplored complexity of the low luminosity XLFs.

A comparison of the low-luminosity behavior of early-type galaxies XLFs with the XLFs of M31 will be pivotal. In particular, the CG XLF of M31 has a distinctive break at 2-5 × $10^{37}$ erg s$^{-1}$ (Kong *et al.* 2002, 2003; Trudolyubov & Priedhorsky 2004). The discovery of a similar break in the E and S0 XLFs would argue for a GC-LMXB connection in these galaxies. The 'outburst peak luminosity – orbital period' correlation (King & Ritter 1998) predicts a break at this luminosity if a large fraction of the sources are ultra-compact neutron star systems. This is intriguing, since the formation of ultra-compact LMXBs is favored in Milky Way GCs (Bildsten & Deloye 2004).

### 3.3.3 Normalization

After shape (breaks and functional 'slopes'), normalization is the remaining parameter of XLFs. Although the early work of Kraft *et al.* (2001) concluded that the number of LMXB per unit optical light in Cen A, M31, M84 and NGC4697 is variable, more recent work on larger samples of galaxies has uncovered strong and convincing trends. Gilfanov (2004) notices a good correlation between the cumulative luminosity above $10^{37}$ erg s$^{-1}$, based on his generalized XLF shape, and the stellar mass of the galaxies. That stellar mass is the main regulator of the number of LMXBs in a galaxy is not surprising, considering that LMXBs are long-lived systems (this connection was also inferred from X-ray-optical/near-IR correlations in bulge-dominated spirals with *Einstein* data; Fabbiano, Gioia & Trinchieri 1988; Fabbiano & Shapley 2002). However, more complexity may be hidden in these correlations. Kim & Fabbiano (2004; see also Humphrey & Buote 2004 for general agreement with this correlation in the case of NGC1332) find a correlation between K-band luminosity (proportional to stellar mass) and integrated LMXB luminosity, but also note that this correlation has more scatter than would be expected in terms of measurement errors. This scatter appears correlated with the GC specific frequency of the galaxies (the number of GC per unit light), suggesting a role of GCs in LMXB evolution (see suggestion of White, Sarazin & Kulkarni 2002). This GC connection will be discussed below.

## 3.4 Association of LMXBs with GCs: the 'Facts'

The association of LMXBs with GCs, and the role of GCs in LMXB formation have been the source of much debate (see reviews by Verbunt and Lewin 2005; Fabbiano & White 2005). The discovery of a significant fraction of LMXBs in GCs in E and S0 galaxies dates from the earliest observation of these populations with *Chandra*. Sarazin, Irwin & Bregman (2000) reported this association in NGC4697 and speculated on the role of GCs in LMXB formation, revisiting the original suggestion of Grindlay (1984) for the evolution of bulge sources in the Milky Way. In virtually all E and S0 galaxies with existing good coverage of GCs, both from the ground, and better from *Hubble*, LMXBs are found in GCs. Below, I summarize the observational results on the association of LMXBs with GCs in early-type galaxies from the large body of papers available in the literature. In Sections 3.5 and 3.6, I will discuss the implications of these results.

3.4.1 Statistics

It appears that in general ~4% of the GCs in a given galaxy are likely to be associated with a LMXB (e.g., NGC1399, Angelini, Loewenstein & Mushotzky 2001; NGC4472, Kundu, Maccarone & Zepf 2002; NGC1553, NGC4365, NGC4649, NGC4697, see Sarazin *et al.* 2003; NGC1339, Humphrey & Buote 2004; M87, Jordan *et al.* 2004; Kim *et al.* 2005). Not surprisingly, as first noticed by Maccarone, Kundu & Zepf (2003), the number of LMXBs associated with GCs varies, depending on the GC specific frequency of the galaxy, which is also a function of the morphological type. Sarazin *et al.* (2003) point to this dependence on the galaxy Hubble type, with the fraction of LMXBs associated with GCs increasing from spiral bulges (MW, M31 ~10-20%), to S0s ~20% (NGC1553, Blanton, Sarazin & Irwin 2001; see also NGC5128, where 30% of the LMXBs are associated with GCs, Minniti *et al.* 2004), E ~50% (NGC4697, Sarazin , Irwin & Bregman 2000; NGC4365, Sivakoff, Sarazin & Irwin 2003; NGC4649, Randall, Sarazin & Irwin 2004; see also NGC4552, with 40% of sources in GCs, Xu *et al.* 2005), cD~70% (in NGC1399, Angelini, Loewenstein & Mushotzky 2001; see also M87, where 62% of the sources are associated with GCs, Jordan *et al.* 2004).

3.4.2 Dependence on LMXB and GC luminosity

In NGC1399 (Angelini, Loewenstein & Mushotzky 2001) the most luminous LMXBs are associated with GCs. No significant luminosity dependence of the LMXB-GC association is instead seen in NGC4472 (Kundu, Maccarone & Zepf 2002) or in the four galaxies studied by Sarazin *et al.* (2003), if anything a week trend in the opposite sense. The reverse is however consistently observed: more luminous GCs are more likely to host a LMXB (Angelini, Loewenstein & Mushotzky 2001, Kundu, Maccarone & Zepf 2002, Sarazin *et al.* 2003; Minniti *et al.* 2004; Xu *et al.* 2005; Kim *et al.* 2005); this trend is also observed in M31 (Trudolyubov & Priedhorsky 2004). Kundu, Maccarone & Zepf (2002) suggest that this effect is just a consequence of the larger number of stars in optically luminous GCs. Sarazin *et al.* (2003) estimate a constant probability per optical luminosity of LMXBs to be found in GCs of ~$2.0 \times 10^{-7}$ for $L_X \geq 3 \times 10^{37}$ erg s$^{-1}$.

### 3.4.3 Dependence on GC color

The probability of a GC hosting a LMXB is not a function of the GC luminosity alone. GC color is also an important variable, as first reported by Angelini, Loewenstein & Mushotzky (2001) in NGC1399 and Kundu, Maccarone & Zepf (2002, see also Maccarone, Kundu & Zepf 2003) in NGC4472, and confirmed by subsequent studies (e.g., Sarazin *et al.* 2003; Jordan *et al.* 2004; Minniti *et al.* 2004; Xu *et al.* 2005; Kim *et al.* 2005). In particular, the GC populations in these galaxies tend to be bi-modal in color (e.g., Zepf & Ashman 1993), and LMXB preferentially are found in red, younger and/or metal rich, clusters (V-I >1.1), rather than in blue, older and/or metal poor, ones. In NGC4472, red GCs are three times more likely to host a LMXB than blue ones (Kundu, Maccarone & Zepf 2002). Similarly, in M87, which has a very rich LMXB population, the fraction of red GC hosting a LMXBs is 5.1% ± 0.7% versus 1.7% ± 0.5% for blue GC (Jordan *et al.* 2004), also a factor of three discrepancy. In M31, an association of LMXBs with red metal rich GC is also reported (Trudolyubov & Priedhorsky 2004). In a sample of six ellipticals yielding 285 LMXB-GC associations (Kim *et al.* 2005), the mean probability for a LMXB-GC association is 5.2%, the probability of a blue GC to host a LMXB is ~2 % for all galaxies except NGC1399 (where it is 5.8%), while that of LMXB-red GC association is generally larger, but varies from a galaxy to another (2.7% - 13%).

### 3.4.4 X-ray colors

In NGC4472, LMXBs associated with blue GCs have been found to have harder 'stacked' X-ray spectra than those in red GCs (Maccarone, Kundu & Zepf 2003). However, a subsequent study with a much larger sample of LMXBs from several E and S0 galaxies does not generalize this result, reporting no significant differences in the X-ray colors of LMXBs associated with either red or blue GCs (Kim *et al.* 2005). Also, no significant differences are found in the X-ray colors of LMXBs in the field or in GCs (Sarazin *et al.* 2003; Kim *et al.* 2005).

### 3.4.5 Spatial distributions of field and GC LMXBs

The radial distributions of LMXBs have also been compared with those of the GCs and of field stellar light, to obtain additional constraints on their formation and evolution. Kundu, Maccarone & Zepf (2002) suggest that in NGC4472 the radial distribution of LMXBs resembles more that of the GCs (more extended) than that of the optical light. However, confusion in the inner radii may affect the detection of LMXBs, artificially depressing their number and resulting in an apparent more extended distribution (see Kim & Fabbiano 2003, fig. 4 Left; Gilfanov 2004).

Other authors instead conclude that overall the LMXB distribution and the stellar light trace each other in E and S0 galaxies (NGC1316, Kim & Fabbiano 2003; NGC1332, Humphrey & Buote 2004). In M87 the radial distributions of field and GC LMXBs are consistent with each other, and within statistics cannot be distinguished from those of the stellar light and GCs, which however differ, with the GC one being more extended (Jordan *et al.* 2004). The spatial distributions of field and GC sources are also consistent within statistics in the sample of Sarazin *et al.* (2003). Kim *et al.* (2005) find that the probability of a GC being associated with a LMXB increases at smaller galactocentric radii; this effect is also echoed in the radial profiles of the LMXB surface number density in this sample. These profiles are consistent within statistics for both GC and field LMXBs and are closer to that of the field stellar surface brightness distribution, which is more centrally peaked, than to the overall GC distribution (Fig. 6).

3.4.6 XLFs of field and GC LMXBs

No significant differences have been found in the XLFs of LMXB in the field and in GCs (Kundu, Maccarone & Zepf 2002; Jordan *et al.* 2004). The co-added XLFs of field and GC LMXBs in six elliptical (Kim *et al.* 2005) are also consistent within the errors, with a similar percentage of high luminosity sources with $L_X > 10^{39}$ erg s$^{-1}$.

This similarity of field and GC XLFs does not extend, however, to the X-ray populations of the Sombrero galaxy (Di Stefano *et al.* 2003) and M31 (from a comparison of the XLFs of bulge and GC sources; Trudoyubov & Priedhorsky 2004). In both cases, the GC XLFs show a more pronounced high luminosity break than the field (bulge) XLFs. In M31 the XLF of GC sources is relatively more prominent at the higher luminosities than that of field LMXBs, in the Sombrero galaxy GC sources dominate the emission in the 1-4 × 10$^{38}$ erg s$^{-1}$ range, but there is a high luminosity tail in the field XLF, which however, could be due to contamination from younger binary system belonging to the disk of this galaxy (see Di Stefano *et al.* 2003).

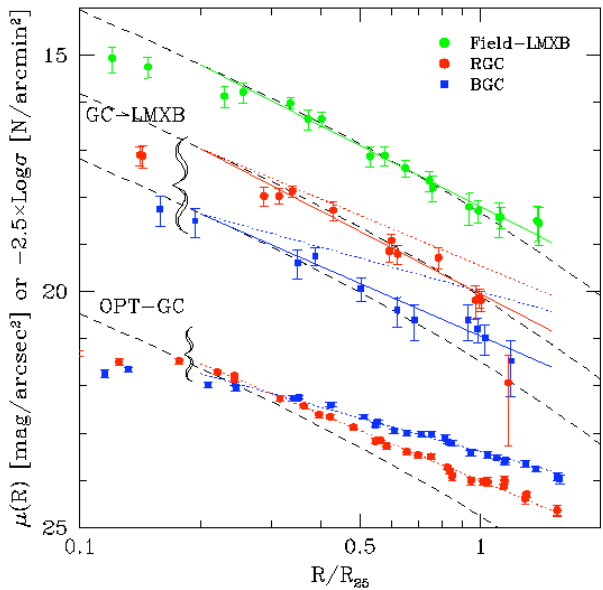

Fig. 6 – Radial distributions of LMXBs in the field (green), in red GCs (red) and in blue GCs (blue), compared with the best-fit GC distributions (dotted, see bottom of the figure), best fit lines (solid) and stellar light distribution (black – dashed). The lower points at small radii may be affected by source confusion in source detection (Kim *et al.* 2005)

3.5 Metallicity and dynamical effects in GC LMXB formation

The preferential association of LMXBs with red clusters could be either an age or a metallicity effect. A correlation between the number density of binaries and the metallicity of GCs was first suggested by Grindlay (1987), who ascribed this effect to a flatter IMF in higher metallicity GCs, resulting in a larger number of neutron stars and thus LMXBs. Kundu *et al.* (2003) argue that metallicity is the main driver, based on the absence of any correlations of LMXB association with different age GC populations in NGC4365. Maccarone, Kundu & Zepf (2004) propose irradiation-induced winds in metal poor stars to speed up evolution and account for the observed smaller numbers of LMXBs in blue GCs. These winds, however, would cause absorption, and thus harder X-ray spectra. Although these authors tentatively reported this spectral effect in NGC4472, other results do not confirm this conclusion (see Section 3.4.4).

While also considering the possibility of irradiation-induced winds (Maccarone, Kundu & Zepf 2004), Jordan *et al.* (2004) re-discuss the IMF-metallicity effect, because the resulting increase in the number of neutron stars agrees with their conclusion that the probability that a GC contain a LMXB is driven by the dynamical properties of the cluster (central density, tidal capture and binary-neutron star exchange rate) as well as metallicity. This conclusion agrees with 3-D hydro-dynamical calculations of the dynamical formation of ultra-compact binaries in GCs, from red giant and neutron star progenitors (Ivanova *et al.* 2005). Kim *et al.* (2005) also invoke dynamical effects to explain the increasing probability of LMXB-GC association at smaller galactocentric radii. These authors suggest that the GCs nearer to the galaxy centers are likely to have a compact core and higher central densities to survive tidal disruption, compared with the GCs at the outskirts, characteristics that would also increase the chance of dynamical LMXB formation.

## 3.6 Constraints on the formation and evolution of LMXBs: field binaries or GC sources?

The formation processes of LMXBs have been debated starting from the discovery of these sources in the Milky Way (see Giacconi 1974). LMXBs may result from the evolution of a primordial binary system, if the binary is not disrupted when the more massive star undergoes collapse and supernova event, or may be formed by capture of a companion by a compact remnant in GCs (see Grindlay 1984; reviews by Verbunt 1993, Verbunt & van den Heuvel 1995). The same scenarios are now being debated for the LMXB populations of E and S0 galaxies. If GCs are the principal (or sole) birthplace, formation kicks or evaporation of the parent cluster have been suggested as an explanation for the field LMXBs in these galaxies (see e.g. Kundu, Maccarone & Zepf 2002).

The correlation (White, Sarazin & Kulkarni 2002) or, more accurately, the second order correlation of the total LMXB luminosity in a galaxy with the GC specific frequency (Kim & Fabbiano 2004), may suggest that GCs are important in the formation of LMXBs. This correlation led White, Sarazin & Kulkarni (2002) to suggest formation in GCs as a universal LMXB formation mechanism in early-type galaxies. The similarity of field and GC LMXB properties (see Section 3.4) has been used in support of a universal GC formation scenario for LMXBs (e.g. Maccarone,

Kundu & Zepf 2003). However, this conclusion is by no means certain or shared by all. Besides uncertainties in the correlations (Kim & Fabbiano 2004), the relationship between the fraction of LMXBs found in GCs and the GC specific frequency (Maccarone, Kundu & Zepf 2003; Sarazin *et al.* 2003) is consistent with the simple relationship expected if field LMXB originate in the field while GC LMXB originate in GCs (Juett 2005; Irwin 2005). This picture would predict different spatial distributions of field and GC LMXB, an effect not seen so far. However, as Juett (2005) notes, the prevalence of LMXB in red (more centrally concentrated) GCs and the effect of SN kicks in the distribution of binaries may make the two distributions less distinguishable.

Piro & Bildsten (2002) and Bildsten & Deloye (2004) compare the observational results with theoretical predictions for the evolution of field and GC binaries. Piro & Bildsten, remark that the large X-ray luminosities of the LMXBs detected in early-type galaxies ($> 10^{37}$ and up to $10^{39}$ erg s$^{-1}$), imply large accretion rates ($> 10^{-9}$ M$_\odot$ yr$^{-1}$). In an old stellar population these sources are likely to be fairly detached binaries that accumulate large accretion disks over time, and undergo transient X-ray events when accretion is triggered by disk instabilities. These transients would have recurrence times greater than 100 yr and outbursts of 1-100 yr duration. They therefore predict that field binaries should be transient, a prediction that is supported by the detection of transients in the NGC5128 LMXB population (Kraft *et al.* 2001). Piro & Bildsten also point out that GC sources tend to have shorter orbital periods and would be persistent sources, reducing the fraction of transients in the LMXB population. Interestingly, Trudolyubov & Priedhosky (2004) report only one recurrent transient in their study of GC sources in M31, although 80% of these sources show some variability; however, they also find six persistent sources in the $10^{38}$ erg s$^{-1}$ luminosity range. Time monitoring observations of the LMXB populations of elliptical galaxies with *Chandra* will be crucial for constraining the fraction and duty cycle of high luminosity transients, and therefore the native field binary component

Bildsten & Deloye (2004) instead look at ultra-compact binaries formed in GCs to explain the bulk of the LMXBs detected in E and S0 galaxies. A motivation for this work is the large probability of finding LMXBs in GCs (per unit optical light, see Section 3.4.2), which makes formation in GCs more efficient than in the field. Ultra-compact binaries would be composed of an evolved low-mass donor star (a white dwarf), filling its Roche lobe, in a 5-10 minutes orbit around a neutron star or a black hole. The entire observable life of such a system is ~$10^7$ yr, much shorter than the age of the galaxies and the GCs, therefore their total number would be indicative of their birth rate. From this consideration Bildsten and Deloye derive a XLF with a functional slope in excellent agreement with the measurements of Kim & Fabbiano (2004) and Gilfanov (2004). Bildsten and Deloye also predict a break at $L_X$ ~$10^{37}$ erg s$^{-1}$ in the XLF, which would correspond to the luminosity below which such a system would be a transient. This prediction can be verified with deep *Chandra* observations. A low luminosity break is suggested by the composite XLF of Gilfanov (2004), which however includes bulges of spiral galaxies. Interestingly, the XLF of GC sources in M31 may have a break at 2-5 × $10^{37}$ erg s$^{-1}$ (Kong *et al.* 2003; Trudolyubov & Priedhorsky 2004).

The nature of the most luminous sources in E and S0 galaxies (those with $L_X$ above the $5 \times 10^{38}$ erg s$^{-1}$ break, Kim & Fabbiano 2004) is the subject of a recent paper by Ivanova & Kalogera (2005). These authors point out that only a small fraction of these luminous sources are associated with GCs (at least in M87, see Jordan *et al.* 2004) and that they are too luminous to be explained easily with accreting neutron star systems that may form in GCs (Kalogera, King & Rasio 2004). With the assumption that these sources are accreting black hole binaries, these authors explore their nature from the point of view of the evolution of field native binaries. In this picture most donor stars would be of low enough mass (<1-1.5 $M_\odot$ given the age of the stellar populations in question), that the binary would be a transient (see Piro & Bildsten 2002), and therefore populate the XLF only when in outburst emitting at the Eddington luminosity; this would happen from main sequence, red giant and white dwarf donors. In this case the XLF is a footprint of the black hole mass spectrum in these stellar populations, which is an important ingredient for linking the massive star progenitors with the resulting black hole. Ivanova & Kalogera derive a differential slope of ~2.5 for the black hole mass spectrum, and an upper black hole mass cut-off at ~20 $M_\odot$, to be consistent with the observed cumulative XLF of Kim & Fabbiano (2004) and Gilfanov (2004). Depending on the magnetic breaking prescription adopted, either red-giant donors or main sequence donors would dominate the source population. A word of caution is order here, since the similar shapes of GC and field LMXB XLFs (Kim *et al.* 2005) suggests that high luminosity black hole sources may also be found in GCs, at odds with theoretical discussions (e.g., Kalogera, King & Rasio 2004).

## 3.7 'Young' Early –type Galaxies and 'Rejuvenation'

Although it is still early to have a solid understanding of the effects of younger stellar age or rejuvenation (e.g. by a merger event or close encounter with a dwarf galaxy) on the X-ray populations of early-type galaxies, there have been some puzzling and somewhat controversial results that suggest that these effects may have some play. This suggestion has been advanced in the case of NGC720 (Jeltema *et al.* 2003), NGC4261 (fig. 7) and NGC4697 (Zezas *et al.* 2003); in these galaxies the source distribution is asymmetric, not following the stellar field light, and at least in NGC720 and NGC4261 the luminosity of the sources exceeds the Eddington limit for a neutron star binary. However, Giordano *et al.* (2005) report the identification of the NGC4261 sources with GCs, undermining the suggestion that they may be linked to a rejuvenation event. Sivakoff, Sarazin & Carlin (2004) report an exceptionally luminous population of 21 sources with $L_X > 2 \times 10^{39}$ erg s$^{-1}$ (in the ULX regime, see Section 6.) in the X-ray bright elliptical NGC1600, twice the number of sources that would be expected from background AGNs, and suggest an XLF slightly flatter than in most ellipticals. In all these cases, however, both cosmic variance affecting the background AGN density and distance uncertainties may play a role.

The opposite behavior to the one just discussed is reported in an X-ray and optical study of the nearby lenticular galaxy NGC5102 (Kraft *et al.* 2005). In this galaxy, where the stellar population is definitely young (<3 × 10$^9$ yr old), and where there is evidence of two recent burst of star formation, a definite lack of X-ray sources is observed. NGC5102 has also a very low specific frequency of GC (~0.4). Kraft *et al.* speculate that the lack of LMXB may be related either to insufficient time for the evolution of a field binary and/or to the lack of GCs.

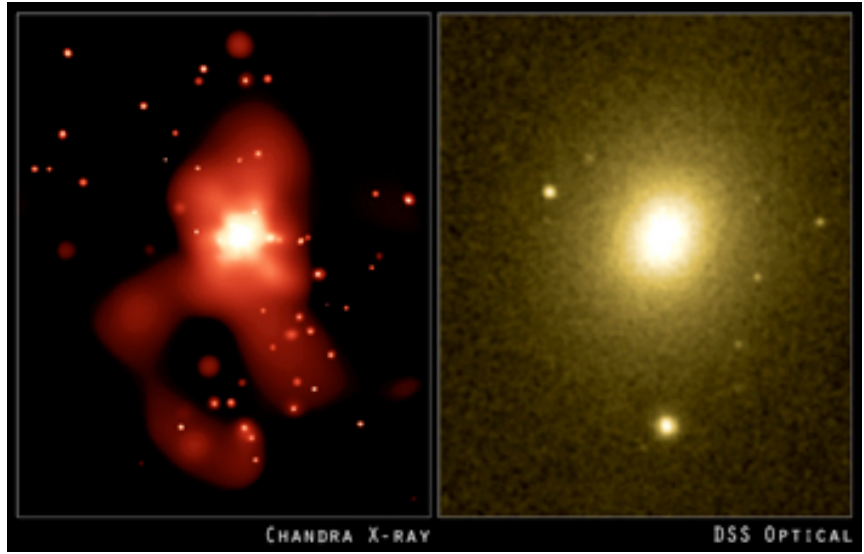

Fig. 7 – Left, Chandra image of NGC4261; Right, optical image (Both images are from http://chandra.harvard.edu/photo/category/galaxies.html; credit NASA/CXC; Zezas *et al.* 2003)

## 4. YOUNG XRB POPULATIONS

The association of luminous X-ray sources (HMXBs, SNRs) with the young stellar population has been known since the dawn of X-ray astronomy (see Giacconi 1974). Luminous HMXBs are expected to dominate the emission of star forming galaxies (Helfand & Moran 2001). These sources, resulting from the evolution of a massive binary system where the more massive star has undergone a supernova event, are short-lived ($\sim 10^{6-7}$ yr), and constitute a marker of recent star formation: their number is likely to be related to the galaxy star formation rate (SFR). This X-ray population – SFR connection was first suggested as a result of the analysis of the sample of normal galaxies observed with *Einstein*, where a strong correlation was found between global X-ray and FIR emission of late-type star-forming galaxies (Fabbiano & Trinchieri 1985; Fabbiano, Gioia & Trinchieri 1988; David, Forman & Jones 1991; Shapley, Fabbiano & Eskridge 2001; Fabbiano & Shapley 2002), and has been confirmed by analyses of *ROSAT* data (Read & Ponman 2001; Lu & Bian 2005).

While HMXBs are likely to dominate the X-ray emission of the most violently star-forming galaxies, they are also expected to be found in more normal spirals (see the Milky Way population; Section 2.), albeit in smaller numbers and mixed with more aged X-ray populations. In bulge dominated spirals, they may constitute only a fraction of the X-ray emitting population, witness the strong correlation between X-ray and H-band luminosity found in these systems (Shapley, Fabbiano & Eskridge 2001; Fabbiano & Shapley 2002). The study of these young populations is then less straightforward than that of LMXBs, because in many cases they must be culled from the complex X-ray source populations of spiral galaxies.

## 4.1 X-ray source populations of spiral and irregular galaxies

The presence of X-ray source populations in Local Group and nearby spiral and late-type galaxies was clearly demonstrated by the early *Einstein* observations (see the 1989 review, Fabbiano 1989; and the *Einstein Catalog and Atlas of Galaxies*, Fabbiano, Kim & Trinchieri 1992). Depending on the prevalence of bulges or of star-forming spiral arms, these X-ray sources may be old systems, such as the LMXBs discussed in Section 3., or younger X-ray emitting sources (HMXBs, dominating at higher luminosities, or SNRs).

Reflecting the complex stellar populations of these galaxies, *Chandra* and *XMM-Newton* studies are reporting clear evidence of complex X-ray source populations. Depending on the exposure time and the distance of the galaxy, from few tens to well over hundred sources have been detected per galaxy. Source variability and spectral analysis have been carried out for the most luminous sources, with both *Chandra* and *XMM-Newton*. In general, the results are reminiscent of the spectral and temporal-spectral behavior of Galactic XRBs, including soft and hard spectral states, and confirm that the luminous sources are indeed accreting binary systems [e.g., M31 (NGC224): Trudolyubov, Borodzin & Priedhorsky 2001, Trudolyubov *et al.* 2002a, Kaaret 2002, Kong *et al.* 2002, Williams *et al.* 2004, Trudolyubov & Priedhorsky 2004, Pietsch, Freyberg & Haberl 2005; M33 (NGC598): Grimm *et al.* 2005, Pietsch *et al.* 2004; NGC1068: Smith & Wilson 2003; NGC1637: Immler *et al.* 2003; NGC2403, Schlegel & Pannuti 2003; M81 (NGC3031): Swartz *et al.* 2003; M108 (NGC3556): Wang, Chavez & Irwin 2003; NGC4449: Miyawaki *et al.* 2004; M104 (NGC4594; Sombrero): Di Stefano *et al.* 2003; M51 (NGC5194/95): Terashima & Wilson 2004; M83 (NGC5236): Soria & Wu 2002; M101 (NGC5457): Pence *et al.* 2001, Jenkins *et al.* 2004, 2005].

In most cases, however, the sources are too faint for detailed analysis and both their position relative to the optical image of the galaxy (e.g., bulge, arms, disk, GCs), their X-ray colors, and in some cases optical counterparts, have been used to aid in the classification. Typically, as demonstrated by Prestwich *et al.* (2003) who applied this method to five galaxies (fig. 3), color-color diagrams can discriminate between harder XRB candidates (with relatively harder HMXBs and softer LMXBs), softer SNR candidates and very soft sources (SSSs, with emission below 1 keV). Similarly, XRBs, SNRs and SSSs are found with *XMM-Newton* X-ray colors in IC342, where most sources are near or on the spiral arms, associating them with the young stellar population (Kong 2003). In M33, the Local Group Scd galaxy with a predominantly young stellar population, *Chandra* and *XMM-Newton* colors, luminosities, and optical counterparts indicate a prevalence of (more luminous) HMXBs and a population of (fainter) SNRs (Pietsch *et al.* 2004; Grimm *et al.* 2005).

In M83 (fig. 2 Left), the X-ray source population can be divided into three groups, based on their spatial, color and luminosity distributions (Soria & Wu 2003): fainter SSSs and soft sources (the latter with no detected emission above 2 keV)), and more luminous and harder XRBs. The soft sources are strongly correlated with current star formation, as indicated by Hα emission in the spiral arms and starburst nucleus, strongly suggesting that they may be SNRs.

*Chandra* X-ray colors (or hardness ratios) were also used to study the X-ray source populations of M100 (NGC4321; Kaaret 2001), M101 (Jenkins *et al.* 2005), NGC1637 (Immler *et al.* 2003), NGC4449 (Summers *et al.* 2004), NGC5494 (the Sombrero galaxy, a Sa, with a predominantly older stellar population; Di Stefano *et al.* 2003), and the star-forming merging pair NGC4038/9, the Antennae galaxy (Fabbiano, Zezas & Murray 2001; Fabbiano *et al.* 2004a), where spectral and flux variability is revealed by color-color and color-luminosity diagrams (Fabbiano *et al.* 2003a, b; Zezas *et al.* 2002a, b, 2005). Colbert *et al.* (2004) employ *Chandra* color diagrams to classify the X-ray source populations in their survey of 32 nearby galaxies of all morphological types, suggesting that hard accreting X-ray pulsars do not dominate the X-ray populations and favoring black hole binaries.

These spectral, photometric and time-variability studies all point to the prevalence of XRB emission at the higher luminosities in the source population, in agreement with what is known from the X-ray observations of the Milky Way and Local Group galaxies (e.g., Helfand & Moran 2001). Comparison of accurate *Chandra* source positions with the stellar field in three nearby starburst galaxies shows that some of these sources experience formation kicks, displacing them from their parent star cluster (Kaaret *et al.* 2004), as observed in the Milky Way. The X-ray luminosity functions, which will be discussed next, can then be considered as reflecting the XRB contribution (LMXBs, and HMXB), with relatively little contribution from the SNRs. This point is demonstrated by the direct comparison of HMXB and SNR luminosity functions in M33 (Grimm *et al.* 2005).

## 4.2 X-ray Luminosity Functions – the XLF of the star-forming population

The XLF of LMXB populations (at high luminosity at least) is well defined by the study of early-type galaxies, which have fairly uniform old stellar populations, and where little if any contamination from a young X-ray source population is expected (Section 3.3). The XLFs of late-type galaxies (spirals and irregulars) are instead the sum of the contributions of different X-ray populations, of different age and metallicity. This complexity was clearly demonstrated by the first detailed studies of nearby galaxies, including a comparison of different stellar fields of M31, with *XMM-Newton* and *Chandra*, yielding different XLFs (e.g., Trudolyubov *et al.* 2002b; Williams *et al.* 2004, Kong *et al.* 2003), and the *Chandra* observations of M81. In this Sab galaxy, the XLF derived from disk sources is flatter than that of the bulge (Tennant *et al.* 2001; fig. 8 Left). In the disk itself, the XLF becomes steeper with increasing distance from the spiral arms. In the arms the XLF is a pure power-law with cumulative slope $-0.48 \pm 0.03$ (Swartz *et al.* 2003), pointing to a larger presence of high luminosity sources in the younger stellar population.

To derive the XLF of HMXB populations, to a first approximation, one must either evaluate the different contributions of older and younger source populations to the XLFs of spiral galaxies, as discussed above in the case of M81, or study galaxies where the star formation activity is so intense as to produce a predominantly young X-ray source population. Both approaches suggest that the HMXB XLF is overall flatter than that of LMXBs, with a cumulative power-law slope of −0.6 to −0.4; in other words, young HMXBs populations contain on average a larger fraction of very luminous sources than the old LMXB populations (see the comparisons of Colbert *et al.* 2004; Kilgard *et al.* 2002; Zezas & Fabbiano 2002; Eracleous *et al.* 2002). These comparisons also show that flatter XLF slopes of about −0.4 to −0.5 are found in intensely star-forming galaxies, such as the merging pair NGC4838/9 (the Antennae galaxies) and M82 (Zezas & Fabbiano 2002; Kilgard *et al.* 2002). In particular, Kilgard *et al.* (2002) find a correlation of the power-law slope with the 60μm luminosity of the galaxy (fig. 8 Right) suggesting that such a flat power-law may describe the XLF of the very young HMXB population. A comparison of dwarf starburst with spiral XLFs (Hartwell *et al.* 2004) is consistent with the above picture: cumulative XLF slopes for spirals are −1.0 to −1.4, slopes for starbursts are lower, −0.4 to −0.8. The connection of the slope with the SFR is demonstrated by comparisons with the 60/100μm ratio, 60μm luminosity and FIR/B ratio.

Grimm, Gilfanov & Sunyaev (2003) took these considerations a significant step further, by comparing the XLFs of 10 star-forming galaxies (taken from the literature), observed with *Chandra* and *XMM-Newton*, with the HMXB luminosity functions of the SMC and the Milky Way. They suggest that there is a universal XLF of star-forming populations, stretching over 4 decades in luminosity ($\sim 4 \times 10^{36} - 10^{40}$ erg s$^{-1}$), with a simple power-law with cumulative slope −0.6. They reach this conclusion by considering that the XLF of star-forming galaxies is likely to be dominated by young and luminous HMXBs, and propose that the SFR per unit stellar mass is the factor responsible for the relative amount of HMXBs in a galaxy; when normalized relative to this quantity, the cumulative XLFs they consider in their study collapse into a single −0.6 power law. The possibility of a 'universal' HMXB XLF is an interesting result, although there are clearly variations in the individual luminosity functions used by Grimm, Gilfanov & Sunyaev (2003), with slopes ranging from ~ −0.4 (the Antennae; Zezas & Fabbiano 2002, where the data were corrected for incompleteness at the low luminosities), to −0.8 (M74-NGC628; Soria & Kong 2002).

A number of XLFs of spiral galaxies have cumulative slopes close to the −0.6 slope of Grimm *et al.* (2003): (IC342, Kong 2003, Bauer, Brandt & Lehmer 2003; NGC5253, Summers *et al.* 2004; NGC4449, Summers *et al.* 2003; NGC2403, Schlegel & Pannuti 2003; NGC6946, Holt *et al.* 2003; NGC1068, Smith & Wilson 2003; NGC2146, Tatsuya *et al.* 2005). However, different and more complex behaviors are also observed, pointing to complexity or evolution of the X-ray source populations. In NGC1637 (Immler *et al.* 2003), the cumulative XLF is reported to follow a power-law of slope −1, for the entire luminosity range covered ($\sim 6 \times 10^{36} - 10^{39}$ erg s$^{-1}$); while a possible break of the XLF ($L_X > 1 \times 10^{37}$ erg s$^{-1}$) is reported, there is no discussion of completeness correction. In NGC2403 (Schlegel & Pannuti 2003), the XLF has cumulative slope −0.6, but this galaxy does not follow the XLF

slope – FIR correlation of Kilgard *et al.* (2002), suggesting it may have stopped forming stars, and we may be observing it while the massive stellar population has evolved, but the HMXBs are still emitting. In NGC6946 (Holt *et al.* 2003), while the cumulative XLF slope is generally consistent with the Grimm, Gilfanov & Sunyaev (2003) conclusions, differences are seen comparing the XLF of the sources in the spiral arms (slope –0.64) with that of sources within 2' of the starburst central region, which is flatter (–0.5). The XLF of NGC5194 (M51, Terashima & Wilson 2004) follows an unbroken power-law with slope –0.9.

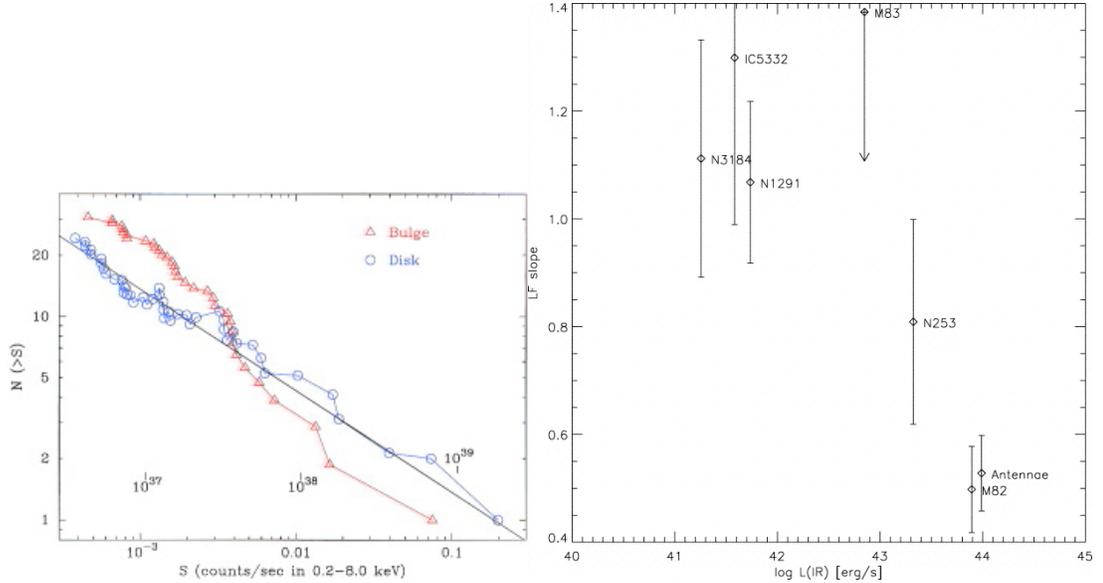

Fig. 8 – Left, bulge and disk XLFs of M81 (Tennant *et al.* 2001). Right, correlation of XLF slope with infrared luminosity (Kilgard *et al.* 2002). Younger stellar populations have flatter XLFs.

In a detailed *Chandra* study of M83, a grand design spiral with a nuclear starburst, Soria & Wu (2003) find that the XLFs of different groups of sources identified by their X-ray colors differ. SSSs, which are found in regions with little or no Hα emission, have a steep XLF, typical of an old population; soft SNR candidates, which tend to be associated with the spiral arms, also have a fairly steep XLF, although it extends to luminosities higher than that of the SSSs; the hard XRB candidates dominate the overall X-ray emission, and therefore the overall XLF. For these sources differences in the XLF are also found, which can also be related to the stellar age. The XLF of the actively star-forming central region is a power law with slope –0.7, instead the XLF of the outer disk has a break at $L_X \sim 8\times10^{37}$ erg s$^{-1}$, and it follows a power law with slope –0.6 below the break, while getting considerably steeper at higher luminosities (–1.6). This type of broken power-law has also been found in the disk of M31 (Williams *et al.* 2004; Shirey *et al.* 2001). In M83, a dip is also seen in the XLF at $\sim 3\times10^{37}$ erg s$^{-1}$, corresponding to 100-300 detected source counts (well above source detection threshold), for sources in the disk and spiral arms where confusion is not a concern. The XLF rises again (towards lower luminosities) after the ~$3\times10^{37}$ erg s$^{-1}$ dip, so incompleteness effects are not likely here. The authors speculate that this complex XLF (fig. 9) may result from an older population of disk sources mixing with a younger (but aging) population of spiral arm sources.

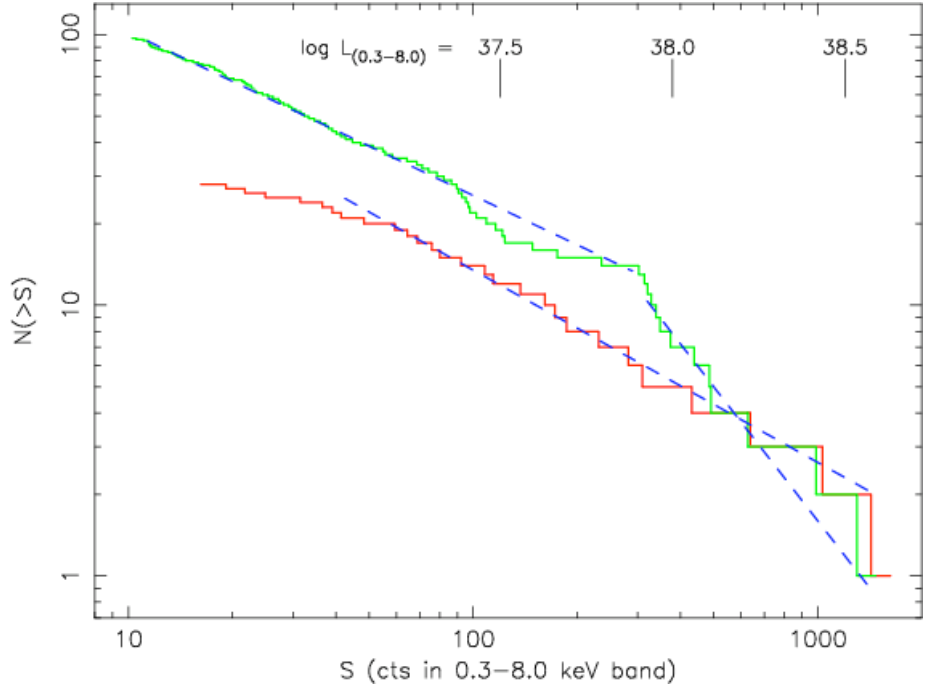

Fig. 9 – XLFs of inner regions (red) and outer disk (green) of M83 (Soria & Wu 2003)

The highest reaches of a star-forming XLF are found in the Cartwheel galaxy (Wolter & Trinchieri 2004), whose detected XRB population is dominated entirely by ULXs. This XLF has a slope consistent with that of Grimm, Gilfanov & Sunyaev (2003), and a large normalization, which suggests a SFR of ~20-25 $M_\odot$ yr$^{-1}$.

The lowest luminosity reaches of the HMXB XLF are explored by Grimm *et al.* (2005) with the *Chandra* survey of M33 (reaching ~ $10^{34}$ erg s$^{-1}$) and by Shtykovskiy and Gilfanov (2005) with the *XMM-Newton* observations of the LMC (reaching ~ $3\times10^{33}$ erg s$^{-1}$). In both galaxies, a large number of the detected sources are background AGNs. In M33, the XLF, corrected for interlopers and incompleteness, is consistent with the HMXB XLF of the Milky Way (Grimm, Gilfanov & Sunyaev 2002). In the LMC, the corrected XLF, with spurious sources removed, and rescaled for the SFR, globally fits the HMXB XLF of Grimm, Gilfanov & Sunyaev (2003) at the high luminosities (~ $10^{37}$ erg s$^{-1}$). The dearth of low-luminosity sources in this XLF leads Shtykovskiy and Gilfanov to suggest a propeller effect (i.e., the magnetic field stopping the accretion flow away from the pulsar surface of the pulsar for relatively low accretion rates). Observing regions of intense local star formation (as indicated by Hα and FIR maxima), where no HMXBs are found, these authors also suggest an age effect, with these star-forming regions being too young for HMXB to have evolved, since HMXB take of order 10 Myr to emerge after the star formation event. Tyler *et al.* (2004) advanced a similar suggestion in their comparison of Hα, mid-IR and *Chandra* images of 12 nearby spiral galaxies.

To conclude, the XLFs of sources in a given galaxy reflects the formation, evolution, and physical properties of the X-ray source populations. These differences are evident for example in different regions of M81 and M83 (see figs. 8 and 9), by comparing elliptical and spiral galaxies and by comparing star-forming galaxies with different SFR. These differences may be related to the aging of the X-ray source population, which will be gradually depleted of luminous young (and short-lived) sources associated with more massive, faster-evolving, donor stars, and also to metallicity effects (Wu 2001; Belczinsky *et al.* 2004). In the future, these X-ray population studies will constitute the baseline against which to compare models of X-ray population synthesis. An early effort towards this end can be found in Belczynski *et al.* (2004).

# 5. SUPER-SOFT SOURCES (SSSs) AND QUASI-SOFT SOURCES (QSSs)

SSSs, as a new class of luminous X-ray sources, were discovered with *ROSAT*. These sources, first found in the Miky Way, M31, the Magellanic Clouds and NGC55, are detected only at energies below 1 keV, and are characterized by spectra that can be fitted with black body temperatures ~15-80 eV (see review by Kahabka & van den Heuvel 1997). Their bolometric luminosities are in the $10^{36}$-$10^{38}$ erg s$^{-1}$ range, and they are believed to be nuclear burning white dwarfs (van den Heuvel *et al.* 1992).

*Chandra* observations have led to the discovery of populations of very soft sources in several galaxies. These newly discovered populations stretch both the spectral and luminosity definition of SSSs, including both slightly harder sources, typically fitted with black-body temperatures ~ 100-300 eV or with a small extra hard component in addition to an SSS spectrum (dubbed QSSs, see Di Stefano & Kong 2003; Di Stefano *et al.* 2004), and super-soft ULXs (in M101, Mukai *et al.* 2003; in the Antennae, Fabbiano *et al.* 2003b; see Section 6.). Variability has been reported in some cases, supporting the idea that these sources are accretion binaries. In M31, a comparison of *Chandra* and *ROSAT* SSSs establishes a variability timescale of several months (Greiner *et al.* 2004); in NGC300 a luminous ($10^{39}$ erg s$^{-1}$) variable SSS is found in *XMM-Newton* data, with a possible 5.4 hr period when in low state (Kong & Di Stefano 2003); the super-soft ULXs in M101 and the Antennae are both highly variable (Mukai *et al.* 2003, 2005; Kong, Di Stefano & Yuan 2004; Fabbiano *et al.* 2003b).

These very soft sources are associated with both old and young stellar populations. They are found in the elliptical galaxies NGC1332 (Humphrey & Buote 2004) and NGC4967 (Di Stefano & Kong 2003, 2004), in the Sombrero galaxy (an Sa; Di Stefano *et al.* 2003), and in a number of spirals: M31 (see Kahabka & van den Heuvel 1997; Kong *et al.* 2002; Di Stefano *et al.* 2004), M81 (Swartz *et al.* 2002), M101 (Pence *et al.* 2001; Di Stefano & Kong 2003, 2004), M83 (Di Stefano & Kong 2003, 2004; Soria & Wu 2003), M51 (Di Stefano & Kong 2003, 2004; Terashima & Wilson 2004), IC342 (Kong 2003), NGC300 (with *XMM-Newton*; Kong & Di Stefano 2003), NGC4449 (Summers *et al.* 2003). Very soft sources are found both in the arms of spiral galaxies, suggesting systems of $10^8$ yr age or younger (see e.g., Di Stefano & Kong 2004), and in the halo and in the bulges, suggesting older counterparts; very soft sources in bulges tend to concentrate preferentially nearer the nuclei (Di Stefano *et al.* 2003; 2004). A QSS is associated with a GC in the Sombrero galaxy (Di Stefano *et al.* 2003). Pietsch *et al.* (2005) report a significant association of SSSs in M31 and M33 with optical novae.

As discussed in several of the above mentioned papers, these results, and the spectral and luminosity regimes discovered with *Chandra* and *XMM-Newton*, strongly suggest that these very soft sources may constitute a heterogeneous population, including both hot white dwarf systems (SSSs), and black hole (or neutron star) binaries (QSSs, super-soft ULXs).

# 6. ULTRA-LUMINOUS X-RAY SOURCES (ULXS)

The most widely used observational definition of ULXs is that of sources detected in the X-ray observing band-pass with luminosities of at least $10^{39}$ erg s$^{-1}$, implying bolometric luminosities clearly in excess of this limit. ULXs (also named intermediate luminosity X-ray objects – IXOs; Colbert & Ptak 2002) were first detected with *Einstein* (Long & Van Speybroeck 1983; see the review Fabbiano 1989), and dubbed super-Eddington sources, because their luminosity was significantly in excess of the Eddington limit of a neutron star (~2 × $10^{38}$ erg s$^{-1}$), suggesting accreting objects with masses of 100 $M_\odot$ or larger. Since these masses exceed those of stellar black holes in binaries (which extends up to ~30 $M_\odot$, Belczynski, Sadowski & Rasio 2003), ULXs could then be a new class of astrophysical objects, possibly unconnected with the evolution of the normal stellar population of a galaxy. They could represent the missing link in the black hole mass distribution, bridging the gap between stellar black holes and the super-massive black holes found in the nuclei of early type galaxies. These 'missing' black holes have been called intermediate mass black holes (IMBH), and could be the remnants of hierarchical merging in the early universe (Madau & Rees 2001), or could be forming in the core collapse of young dense stellar clusters (e.g. Miller & Hamilton 2002). On the other hand, ULXs could just represent a particular high-accretion stage of X-ray binaries, possibly with a stellar black hole accretor (King *et al.* 2001), or even be powered by relativistic jets, as in microquasars (Koerding, Falke & Markoff 2002).

Given these exciting and diverse possibilities it is not surprising that ULXs have generated a lot of both observational and theoretical work. An in-depth discussion of all this work is beyond the scope of the present review. Among the recent reviews on ULXs, presenting different points of view, are those of Fabbiano (2004), Miller and Colbert (2004), Mushotzky (2004) and Fabbiano and White (2005). Two recent short articles in *Nature* and *Science* (McCrady 2004 and Fabbiano 2005) are also useful examples of different perspectives on this subject: McCrady argues for the IMBH interpretation of ULXs, Fabbiano instead concludes that although a few very luminous ULXs are strong candidates for IMBHs, the majority may be just the upper luminosity end of the normal stellar population. Here, I will discuss the main points of the current debate on ULXs, as they pertain to the discourse on X-ray populations, quoting only recent and representative work.

## 6.1 Association of ULXs with active star-forming stellar populations

From a population point of view it is useful to see where we find ULXs. The heightened recent interest in ULXs has spurred a number of studies that have sought to take a systematic view of these sources. These include both works using the *Chandra* data archive, and also those revisiting the *ROSAT* data and the literature. From a mini survey of 13 galaxies observed with *Chandra*, including both ellipticals and spirals, Humphrey *et al.* (2003) suggested a star-formation connection on the basis of a strong correlation of the number of ULXs per galaxy with the 60μm emission and a lack of correlation with galaxy mass. Swartz *et al.* (2004) published spectra, variability and positions for 154 ULXs in 82 galaxies from the *Chandra* ACIS archive, confirming their association with young stellar populations, especially those of merging and colliding galaxies. This conclusion is in agreement with that of Grimm, Gilfanov and Sunyaev (2003), based on a comparison of XLFs of star forming galaxies (see Section 4.2). The strong connection of ULXs with star formation is also demonstrated by the analysis of a catalog of 106 ULXs derived from the *ROSAT* HRI observations of 313 galaxies (Liu & Bregman 2005). Liu & Mirabel (2005) instead compile a catalog of 229 ULXs from the literature, together with optical, IR, and radio counterparts, when available; they observe that the most luminous ULXs (those with $L_X > 10^{40}$ erg s$^{-1}$), which are the most promising candidates for IMBHs, can be found in either intensely star-forming galaxies or in the halo of ellipticals (the latter however are likely to be background QSOs, see below). The association of ULXs with high SFR galaxies is exemplified by the discovery of 14 of these sources in the Antennae galaxies (fig. 10), the prototype galaxy merger.

As discussed in Section 3.3.1, the XLFs of E and S0 galaxies are rather steep, i.e. the number of very luminous sources in these LMXB populations is relatively small, especially in comparison with star-forming galaxies; however, sources with luminosities in excess of $10^{39}$ erg s$^{-1}$ exist (see an earlier discussion of this topic in Fabbiano and White 2005; see also Section 3.7).

Several authors have considered the statistical association of ULXs with early–type galaxies (E and S0s, old stellar populations). Swartz *et al.* (2004) find that the number of ULXs in early type galaxies scales with galaxy mass and can be explained with the high luminosity end of the XLF (see Gilfanov 2004 and discussion in Section 3.3.1). They also point out that ULX detections in early-type galaxies are significantly contaminated by background AGNs, in agreement with the statistical work of Ptak & Colbert (2004) and Colbert & Ptak (2002), based on the *ROSAT* HRI (5'' resolution) observations of galaxies. Irwin, Bregman & Athey (2004) find that sources in the 1-2 × $10^{39}$ erg s$^{-1}$ luminosity range are statistically associated with the galaxies and have spectra consistent with those of Galactic black hole binaries (Irwin, Athey & Bregman 2003). Sources more luminous than 2 x $10^{39}$ erg s$^{-1}$ (if placed at the distance of the associated galaxy) are instead consistent with the expected number and spatial distribution of background AGNs.

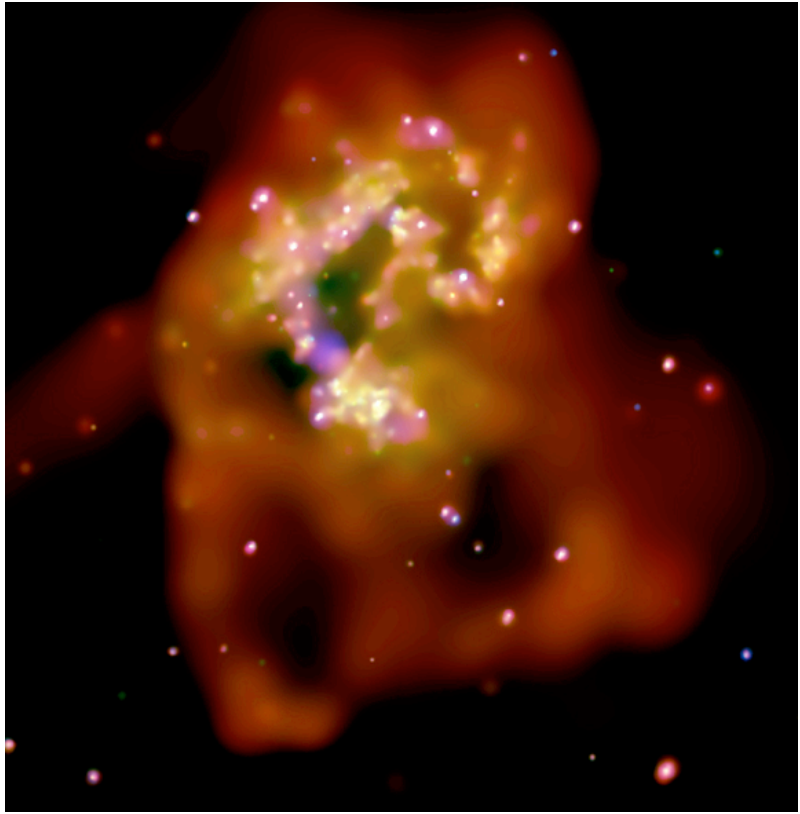

Fig.10 - *Chandra* ACIS image of the Antennae from 2 yr monitoring (4.8' side box; from the web page http://chandra.harvard.edu/photo/category/galaxies.html; credit NASA/CXC; Fabbiano *et al.* 2004a).

This growing body of results demonstrates that ULXs are associated with the star-forming population. The presence of ULXs in early-type galaxies has been debated, but there is no strong statistical evidence for the existence of a population of sources with $L_X$ > 2×$10^{39}$ erg s$^{-1}$ in these galaxies. In the following I will only discuss ULXs in star-forming galaxies.

## 6.2 Spectra and variability from *Chandra* and *XMM-Newton* observations

*Chandra* and *XMM-Newton* work has confirmed that ULXs are compact accreting sources, building on the more limited observations of nearby ULXs with *ASCA* (Makishima *et al.* 2000, Kubota *et al.* 2001). Flux-color transitions have been observed in a mumber of ULXs, suggesting the presence of an accretion disk (in the Antennae, Fabbiano, Zezas & Murray 2001; Fabbiano *et al.* 2003a, b, 2004a; Zezas *et al.* 2005; M101, Jenkins *et al.* 2004; NGC7714, Soria & Motch 2004; M33, LaParola *et al.* 2003; Dubus, Charles & Long 2004; Foschini *et al.* 2004; Ho II X-1, Dewangan *et al.* 2004; and a sample of 5 ULXs in different galaxies monitored with *Chandra*, Roberts *et al.* 2004). Some of these spectra and colors are consistent with or reminiscent of those of black hole binaries (see above refs. and Colbert *et al.* 2004; Liu *et al.* 2005). A recent spectral survey with *XMM-Newton* finds different spectral types, suggesting either spectral variability or a complex source population (Feng & Kaaret 2005).

Shorter-term variability is also consistent with the presence of X-ray binaries and accretion disks. In particular a ULX in NGC253 has recently been shown to be a recurrent transient (Bauer & Pietch 2005). Moreover, features in the power density spectra have been used to constrain the mass of the accreting black hole (Strohmayer & Mushotzky 2003; Soria *et al.* 2004). In the very luminous M82 ULX ($L_x > 10^{40}$ erg s$^{-1}$, $L_{bol} \sim 10^{41}$ erg s$^{-1}$), which is the most compelling IMBH candidate, Strohmayer & Mushotzky (2003) detect a 55mHz QPO, also confirmed by Fiorito & Titarchuck (2004).

The most statistically significant spectra are those obtained with *XMM-Newton* in nearby very bright ULXs where confusion with other unresolved emission in the detection area is not severe. In several cases, these spectra are best fitted by a composite model including a power-law and a very soft accretion disk component (there are exceptions, e.g. for the $10^{41}$ erg s$^{-1}$ ULX in NGC2276, where a multicolored disk model is preferred, Davis & Mushotzky 2004). A very soft component was first reported by Miller *et al.* (2003) for the ULX NGC1313 X-2, with a the temperature of ~150 keV, consistent with the expected temperature of an accretion disk surrounding an IMBH of nearly 1000 M$_\odot$ (but see a more recent estimate of 100 M$_\odot$, Zampieri *et al.* 2004). Similar soft components were found in other ULXs (Miller, Fabian & Miller 2004a; Miller *et al.* 2004; Jenkins *et al.* 2005; Roberts *et al.* 2005).

Unfortunately, these results are not the smoking gun that one may have hoped for to conclusively demonstrate the presence of IMBHs in ULXs. Two other models have been proposed that fit equally well the data, but are consistent with normal stellar black hole masses. One is the slim disk model (e.g., Watarai *et al.* 2005, Ebisawa *et al.* 2004, advanced to explain the emission of an accretion disk in a high accretion mode; see Foschini *et al.* 2005; Roberts *et al.* 2005). The second model is a physical Comptonized disk model (Kubota, Makishima & Done 2004). Although both models are significantly more complex than the power-law + soft-component model, nature can easily be wicked, and the models are physically motivated. The controversy is raging, given the tantalizing possibility of proving the discovery of IMBHs (see Miller, Fabian & Miller 2004b; Fabian, Ross & Miller 2004; Wang *et al.* 2004).

The recurrent variable very soft ULX in M101 provides an excellent case study, to illustrate the difficulty of reaching a firm conclusion on the presence of an IMBH. Given their very soft spectra, SSSs and QSSs in the ULX luminosity range are IMBH candidates (in Sombrero, Di Stefano *et al.* 2003; M101, Mukai *et al.* 2003, 2005; Kong, Di Stefano & Yuan 2004; the Antennae, Fabbiano *et al.* 2003b). These sources are too luminous to be explained in terms of hot white dwarfs, unless the emission is beamed, which is unlikely (e.g., Fabbiano *et al.* 2003b).

The expanding black hole photosphere of a stellar black hole was first suggested to explain the M101 very soft ULX (Mukai *et al.* 2003), but the subsequent detection of a hard power-law component and low/hard – high/soft spectral variability pointed to a Comptonized accretion disk in a black hole binary (Kong, Di Stefano & Yuan 2004; Mukai *et al.* 2005); but what kind of black hole?

Based on the *XMM-Newton* spectrum, which can be fitted with an absorbed black-body, and implies outburst luminosities in the $10^{41}$ erg s$^{-1}$ range, Kong, Di Stefano & Yuan (2004) advanced the IMBH candidacy. Mukai *et al.* (2005), instead, argue for a 20-40 $M_\odot$ stellar black hole counterpart. Their main point is that the high $L_X$ derived in the previous study results from the adoption of an emission model with a considerable amount of line of sight absorption; the colors of the optical counterpart, instead, are consistent with very little absorption; moreover, if the obscuring material were close to the black hole it would be most likely ionized (warm absorber). Adopting an accretion disk plus emission line model, Mukai *et al.* (2005) obtain luminosities in the $10^{39}$ erg s$^{-1}$ range. They also use the variability power density spectrum of the source to constrain the emission state, and with the luminosity, the mass of the black hole. This is another example where a considerable amount of ambiguity exists in the choice of the X-ray spectral model, and X-ray spectra alone may not give the conclusive answer. The luminous optical counterpart makes this source an obvious candidate for future studies aimed at obtaining the mass function of the system.

## 6.3 Counterparts at other wavelengths

As shown by the example at the end of Section 6.2, identification of ULXs may be crucial for understanding the nature of these sources. Three main classes of counterparts have been discussed in the literature: stellar counterparts; ionized or molecular nebulae; and radio sources. Stellar counterparts tend to be early type stars, pointing to HMXBs (Soria *et al.* 2005; Liu, Bregman & Seitzer 2004; Kaaret, Ward & Zezas 2004; Zampieri *et al.* 2004; see also Fabbiano & White 2005 for earlier references); while these counterparts are consistent with the high accretion rate model of ULXs (e.g. King *et al.* 2001; Rappaport, Podsiadlowski & Pfahl 2005), these results do not really constrain the nature of the compact object. Nebular counterparts suggest isotropic emission in some cases, and so a truly large $L_X$, arguing against a substantial amount of beaming and pointing to fairly massive black holes (Roberts *et al.* 2003; Pakull & Mirioni 2003; Kaaret, Ward & Zezas 2004); radio counterparts have been alternately been found consistent with either beamed sources or IMBHs (Kaaret *et al.* 2003; Neff, Ulvestad & Campion 2003; Miller, Mushotzy & Neff 2005; Koerding, Colbert & Falke 2005). Optical variability studies of the stellar counterparts are needed to firmly measure the mass of the system. The new generation of large-area, high-resolution, optical telescopes are likely to solve the nature of these ULXs.

In more distant systems, like the Antennae (D~19 Mpc; Zezas *et al.* 2002a, 2002b,

2005) or the Cartwheel galaxies (D~122 Mpc; Gao *et al.* 2003; Wolter & Trinchieri 2004; King 2004), where spectacular populations of ULXs are detected, individual stellar counterparts cannot be detected. However, comparison with the optical emission field also provides very interesting results. In the Antennae, ULXs tend not to coincide with young star clusters, suggesting that either the system has been subject to a SN formation kick to eject it from its birthplace (thus implying a normal HMXB with a stellar mass black hole; Zezas *et al.* 2002b; Sepinsky, Kalogera & Belczynski 2005), or that the parent cluster has evaporated, in the core collapse model of IMBH formation (e.g., Portegies Zwart *et al.* 2004). However, a recent paper suggests that some of these displacements may be reduced with better astrometric corrections (Clark *et al.* 2005). In the Cartwheel, the ULXs are associated with the most recent expanding star-formation ring, setting strong constraints to the IMBH hypothesis and favoring the high accretion HMXB scenario (King 2004). It must be said, however, that given the distance of this galaxy, and the lack of time monitoring, it cannot be excluded that the ULXs may represent clumps of unresolved sources.

6.3.1 Association with QSOs

Recent optical work of identification of candidate ULXs in both early and late–type galaxies has found some higher red-shift galaxy and QSO counterparts (Masetti *et al.* 2003; Arp, Gutierrez & Lopez-Corredoira 2004; Gutierrez & Lopez-Correidora 2005; Burbidge *et al.* 2004; Galianni *et al.* 2005; Arp & Burbidge 2005; Clark *et al.* 2005; possibly Ghosh et al 2005). Some of these sources are in the halo of early-type galaxies, in agreement with statistical expectations of chance coincidences (see Section 6.1), but QSO identifications in star-forming galaxies have also been reported. While at this point these identifications are still few and consistent (within small number statistics) with chance coincidences with background AGN, some of the above authors (Arp, M. Burbidge, G. Burbidge and collaborators) have raised the hypothesis of a physical connection between the QSO and the parent galaxy; clearly this possibility cannot be extended to the entire body of ULXs, given the results of other identification campaigns (see above).

6.4 Models of ULX formation and evolution

I will summarize here some of the more recent theoretical work on ULXs, and refer the reader to the reviews cited earlier for details on earlier work. As I've already noted, the two principal lines of thought are: (1) most ULXs are IMBHs; (2) most ULXs are luminous X-ray sources of 'normal' stellar origin and the IMBH explanation should be sought only for the ULXs with $L_{bol} > 10^{41}$ erg s$^{-1}$ (the M82 ULX, Matsumoto *et al.* 2001, Kaaret *et al.* 2001; the NGC2276 ULX, Davis & Mushotzky 2004; the most luminous ULX in the Cartwheel galaxy, Gao *et al.* 2003, Wolter & Trinchieri 2004; and the variable ULX in NGC7714, Soria & Motch 2004).

    The stellar evolution camp was originally stimulated by the abundance of ULXs in star-forming galaxies (King *et al.* 2001; King 2004) and by the apparently universal

shape of the XLF of the star-forming population (Grimm, Gilfanov & Sunyaev 2003; see Section 4.2). The variability and spectra of these systems (see Section 6.2) point to accretion binaries. In this paradigm, the problem is to explain the observed luminosities. Both relativistic (Koerding, Falke & Markoff 2002) and non-relativistic beaming (King *et al.* 2001), and super-Eddington accretion disks (Begelman 2002; both spectra and observed variability patterns can be explained, M. Belgelman 2005, private communication) have been suggested, as a way to explain the source luminosities inferred from the observations. With the exception of relativistic beaming, these mechanisms can account for a factor of 10 enhancement of the luminosity above the Eddington value. If black-hole masses of a few tens solar masses exist (Belczynski, Sadowski & Rasio 2004), most or all the ULXs could be explained this way. For example, Rappaport, Podsiadlowski & Pfahl (2005) have combined binary evolution models and binary population synthesis, finding that for donors with M≥10 $M_\odot$, accretion binaries can explain the ULXs, with modest violation of the Eddington limit.

The IMBH camp has generated a larger volume of papers. IMBHs may be remnants of collapse in the early universe (e.g., Van der Marel 2004; Islam, Taylor & Silk 2004), or may result from the collapse of dense stellar clusters (e.g., Gurcan, Freitag & Rasio 2004; Portegies Zwart *et al.* 2004). In the cosmological remnant options, one would expect IMBHs to be particularly abundant in the more massive elliptical galaxies, contrary to the observed association with star-forming galaxies (Zezas & Fabbiano 2002). However, IMBHs would not be visible unless they are fueled, and fuel is more readily available in star-forming galaxies, in the form of dense molecular clouds (Krolik 2004). Accretion from a binary companion is an efficient way of fueling an IMBH, and consequently a number of papers have explored the formation of such binaries via tidal capture in globular clusters. In this picture, the ULX may not be still associated with the parent cluster because of cluster evaporation (Li 2004; Portegies Zwart, Dewi & Maccarone 2004; Hopman, Portegies Zwart & Alexander 2004; Portegies Zwart *et al.* 2004). A twist to the cosmological hypothesis is given by the suggestion that the very luminous ULXs, with $L_{bol} > 10^{41}$ erg s$^{-1}$ such as the M82 ULX, may be the nuclei of satellite galaxies, switching on in presence of abundant fuel (King & Dehnen 2005).

Some of this work has resulted in predictions that can be directly compared with the data. In particular, the slope and normalization of the high-luminosity XLFs of star-forming galaxies have been reproduced in both IMBH (Islam, Taylor & Silk 2004; Krolik 2004) and jet models (Koerding, Colbert & Falke 2004). Zezas & Fabbiano (2002) discuss the effect of either a beamed population of ULXs or a population of IMBH in the context of the XLF of the Antennae. Gilfanov, Grimm & Sunyaev (2004b) predict a change of slope in the $L_X$-SFR relation, where $L_X$ is the total X-ray luminosity of a galaxy, if a new population of IMBHs is present at the higher luminosities (see Section 7.).

Other properties have also been investigated. Besides the X-ray spectral distribution (see Section 6.2), the presence of radio emission from IMBHs has been discussed (e.g., Maccarone 2004), and variability-based tests have been proposed for discriminating between IMBH and stellar black hole binaries, including QPO frequency (that may be a function of back-hole mass, Abramowicz *et al.* 2004) and long-term transient behavior (expected from IMBH binaries, while thermal-timescale mass transfer onto stellar black holes would produce stable disks, Kalogera *et al.* 2004). These tests require long term monitoring of ULXs and future larger X-ray telescopes.

## 7.0 X-RAY EMISSION AND GALAXY EVOLUTION

*Chandra* observations of galaxies at high redshifts (z > 0.1), either from identification of *Chandra* deep survey sources or from stacking analysis of distant galaxy fields have been reviewed recently in the literature (see Fabbiano & White 2005; Brandt & Hasinger 2005) and will not be discussed in detail here. In summary, the emission from normal galaxies becomes an increasingly greater component of the X-ray emission at the deepest X-ray counts (Bauer *et al.* 2004; Ranalli, Comastri & Setti 2005); moreover, the hard X-ray emission is a direct diagnostic of star formation, as demonstrated by the good FIR-X-ray correlations and by the work on the XLFs of star-forming galactic populations discussed earlier in this review (Fabbiano & Shapley 2002; Colbert *et al.* 2004; Ranalli, Comastri & Setti 2003; Grimm, Gilfanov & Sunyaev 2003, Gilfanov, Grimm & Sunyaev 2004a). It is clear that the studies of the global properties and luminosity functions of galaxies at different redshifts can give information in this area, and this work is beginning to gather momentum (e.g. Georgakakis *et al.* 2003; Hornschemeier *et al.* 2005; Norman *et al.* 2004; Ranalli, Comastri & Setti 2005), given the availability of *XMM-Newton* surveys of the nearby Universe and the increasingly deep *Chandra* surveys.

Enhanced star formation early in the life of a galaxy is expected to produce enhancements in its X-ray emission at different epochs, related to the formation and evolution of HMXB and LMXB populations (Ghosh & White 2001). Lehmer *et al.* (2005) report such an effect in their stacking analysis of Lyman break galaxies in the *HST GOODS* fields covered by deep (1-2 Ms exposure) *Chandra* fields (fig. 11).

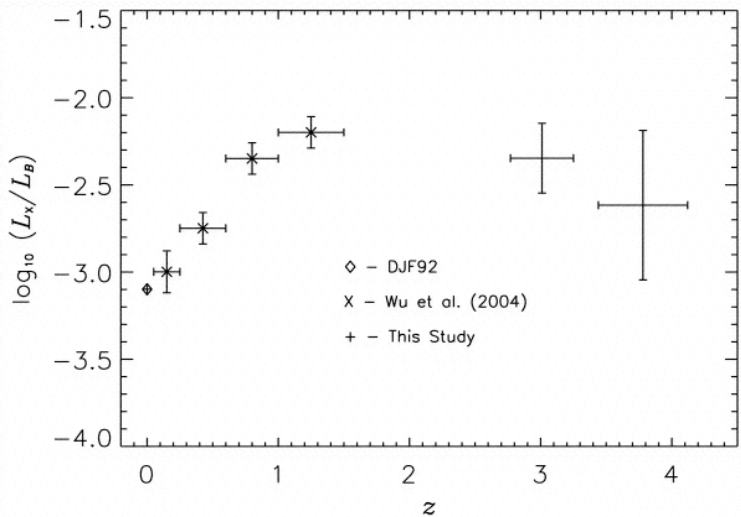

Fig. 11 – Evolution of the X-ray to optical ratio of galaxies with redshift, peaking at z~1.5 – 3 (Lehmer *et al.* 2005)

Conversely, if the SFR is independently known, the relation between the integrated luminosity of galaxies and the SFR can be used to measure the maximum luminosity of a HMXB and the presence of a very high luminosity IMBH population not related to stellar sources (Gilfanov, Grimm & Sunyaev 2004b). These authors, based on the XLF – SFR connection (Grimm, Gilfanov & Sunyaev 2003), explore the statistical properties of a population of discrete sources and demonstrate that a break is expected in the relation between the total X-ray luminosity of the galaxies and the SFR between non-linear (low luminosity – low SFR regime) and linear (high luminosity – high SFR), dependent on the high luminosity cut-off of the XRB population. Comparing the local galaxy sample with the *Hubble* Field North galaxies, they suggest a cut-off luminosity ~5 x $10^{40}$ erg s$^{-1}$ for HMXBs. They also suggest that a population of very luminous IMBHs ($L_X > 10^{40}$ erg s$^{-1}$) would reveal itself with a steeper $L_X$ – SFR relation at higher luminosities and star formation regimes, where here $L_X$ is the total integrated luminosity of the galaxies.

It is clear that very deep high resolution X-ray observations, combined with the understanding of X-ray source populations resulting from *Chandra* observations of nearby galaxies will produce unique results for both the understanding of galaxy evolution and of binary stars formation and evolution.